\documentclass{article}

\usepackage{microtype}
\usepackage{graphicx}
\usepackage{subcaption}
\usepackage{booktabs} 

\usepackage{hyperref}

\usepackage[preprint]{icml2026}

\usepackage{amsmath}
\usepackage{amssymb}
\usepackage{mathtools}
\usepackage{amsthm}

\usepackage[capitalize,noabbrev]{cleveref}

\usepackage[table]{xcolor}         
\definecolor{tablegray}{gray}{0.92}
\definecolor{tablelightgray}{gray}{0.95}
\definecolor{tabledarkgray}{gray}{0.87}

\definecolor{stdgray}{gray}{0.5}
\definecolor{mlgray}{gray}{0.95}
\definecolor{mtgray}{gray}{0.85}

\usepackage{multirow}
\usepackage{subcaption}
\usepackage{graphicx}
\usepackage{enumerate}
\usepackage{enumitem}
\usepackage[mathscr]{euscript}

\usepackage{booktabs}

\usepackage{multirow}
\usepackage{tikz}

\usepackage{amsfonts}
\DeclareMathOperator*{\argmax}{arg\,max}

\usepackage{xspace}

\usepackage{xspace}

\newcommand{\shortparagraph}[1]{\noindent\textbf{#1}\quad}

\newcommand{\method}{AG-CLIP\xspace }

\newcommand{\xmark}{%
\tikz[scale=0.23] {
    \draw[line width=0.7,line cap=round] (0,0) to [bend left=6] (1,1);
    \draw[line width=0.7,line cap=round] (0.2,0.95) to [bend right=3] (0.8,0.05);
}}

\hypersetup{
    colorlinks,
    linkcolor={red!80!black},
    citecolor={green!80!black},
    urlcolor={blue!80!black}
}

\newcommand*{\ie}{i.e.,\@\xspace}

\let\cite\citep

\theoremstyle{plain}

\theoremstyle{definition}

\theoremstyle{remark}

\icmltitlerunning{Bioacoustic Geolocation: Species Sounds as Geographic Signals}

\begin{document}

\twocolumn[
  \icmltitle{Bioacoustic Geolocation: Species Sounds as Geographic Signals}




  \begin{icmlauthorlist}
    \icmlauthor{Mustafa Chasmai}{umass}
    \icmlauthor{Wuao Liu}{umass}
    \icmlauthor{Subhransu Maji}{umass}
    \icmlauthor{Grant Van Horn}{umass}
  \end{icmlauthorlist}

  \icmlaffiliation{umass}{University of Massachusetts, Amherst}

  \icmlcorrespondingauthor{Mustafa Chasmai}{mchasmai@umass.edu}
  
  \icmlkeywords{Machine Learning, ICML}

  \vskip 0.1in
\centering
\centerline{\includegraphics[width=0.98\textwidth]{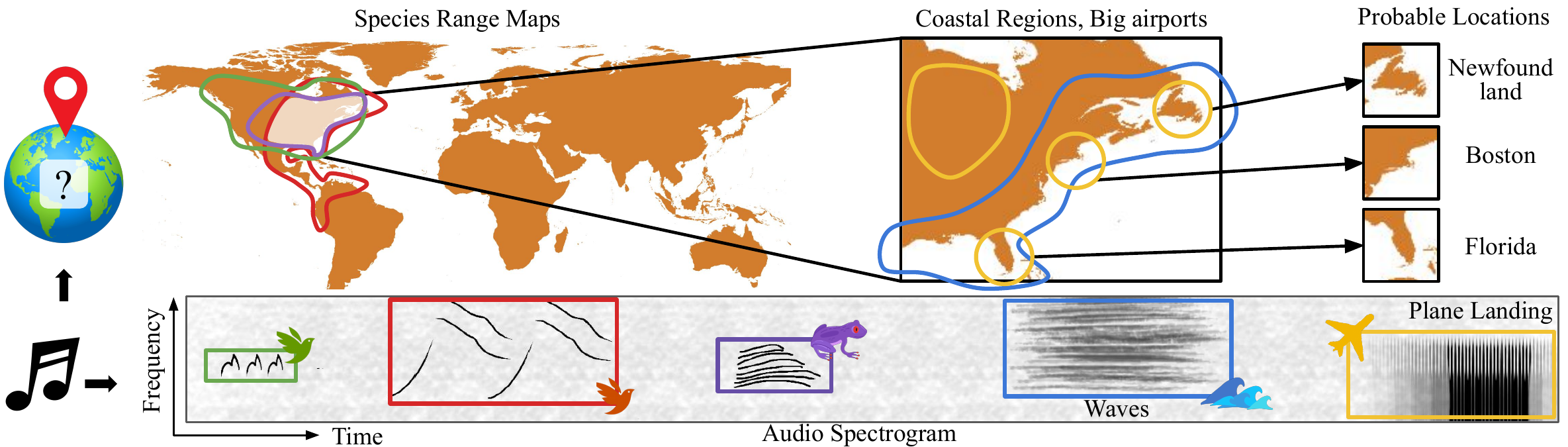}}
\captionof{figure}{\textbf{Intuition for Audio Geolocation.} 
    If a model were to recognize different species' sounds, the intersection of their geographic range maps could be used to estimate the location, with environmental and anthropogenic sounds offering further geospatial calibration. 
     While our models do not explicitly detect these sounds, we hypothesize that they leverage similar signals \textit{implicitly} to perform geolocation. 
    }
    \label{fig:overview}

  \vskip 0.3in
]

\printAffiliationsAndNotice{}  

\begin{abstract}
Can we determine someone’s geographic location solely from the sounds they hear? 
Are acoustic signals enough to localize within a country, state, or even city?  
In this work, we tackle the challenge of global-scale audio geolocation, with a particular focus on wildlife and natural sounds. 
We posit that bioacoustic signals contain informative geolocation cues because of well-defined geographic ranges of species. To test this hypothesis, we benchmark image geolocation and soundscape mapping methods, design oracles and species-centric baselines, and propose a hybrid approach that combines species range prediction with retrieval-based geolocation.
We further ask whether geolocation improves with species-diverse recordings and spatiotemporal aggregation across neighboring samples.
Finally, we extend our study to multimodal geolocation with case studies from movies that combine both audio and visual content.
Our results highlight the potential of incorporating bioacoustic signals into geospatial tasks,  motivating future work on species recognition and audio geolocation.
\end{abstract}    
\section{Introduction}
\label{sec:intro}

Image geolocation has been extensively studied in the computer vision community, with methods exploring classification~\cite{weyand2016planet, vo2017revisiting, seo2018cplanet, berton2022rethinking, pramanick2022world, clark2023we}, contrastive learning~\cite{vivanco2024geoclip, klemmer2023satclip} and multimodal representations~\cite{zhu2021vigor, yang2021cross, zhu2022transgeo}. 
The best models can place 40\% of images within 25 km of their actual location, rivaling top human geoguessers~\cite{haas2024pigeon}.
%
In this work, we shift focus to predicting the recording location of audio, addressing the unique challenges of global audio geolocation.

Image geolocation relies on visual cues such as famous landmarks, landscape types, or architectural styles to infer location from a random image. By contrast, cues tied to a geographic location are less salient in audio.
Early studies~\cite{pokorny2019sound, kumar2016audio} focus on urban sounds like jackhammers and traffic noise as signals to geolocate.  
%
%
We hypothesize that wildlife sounds offer another promising signal. 
Each species has a defined geographic range, and identifying audible species in a recording can help constrain the possible locations.
Intuitively, by intersecting range maps of detected species, we can significantly narrow down the search area (Fig.~\ref{fig:overview}). 
If the detected species have large ranges, then the search area might still involve hundreds or thousands of square kilometers. 
However, if any of the detected species has a restricted range, the plausible search space might be narrowed down to tens of kilometers. 


In this work, we study bioacoustic geolocation using iNatSounds~\cite{chasmai2024inaturalist} and XCDC, a newly curated collection of 
species-rich dawn chorus recordings from \citet{xeno}.
Our experiments reveal that accurate geolocation is possible if we know \emph{all} species present in a location, and performance improves as species diversity increases. 
However, achieving high recall in identifying audible species presents several challenges.
Short audio recordings may fail to capture a representative set of vocalizing species, motivating approaches that aggregate information across multiple recordings within spatiotemporal neighborhoods. 
Even when species are recorded, low signal-to-noise ratios and fine-grained confusions can further limit recall.
Rare species, in particular, have distinctive geographic footprints that are valuable for geolocation, yet detecting them is especially difficult.
These challenges are further exacerbated at the global scale, warranting a dedicated study of geolocation as a task distinct from species identification.

Audio geolocation has numerous potential applications. Coarse geographic context inferred from audio can enable apps such as Merlin Sound ID to constrain the set of plausible species in areas without cell coverage or GPS availability. This geographic context could also be valuable for tracking invasive species or informing conservation planning in data-scarce regions. Soundscapes inherently capture key aspects of a habitat, and shifts in predicted locations of a region over time can be used to monitor ecological change without expensive species labeling. Beyond ecology, audio geolocation can also be applied for digital forensics (Sec.~\ref{sec:geo_forensics}), search and rescue (localizing distress calls), and data privacy (understanding and obfuscating prominent location cues). We see our work as laying the foundation for future efforts and a first step towards ever more diverse applications of the challenging problem of audio geolocation.
Trained models and code are available at \href{https://github.com/cvl-umass/audio-geolocation}{cvl-umass/audio-geolocation}.



We summarize our main contributions as follows:
\begin{enumerate}
[leftmargin=0.04\textwidth, nosep]
\itemsep0em 
    \item 
    To the best of our knowledge, this is the first study on global-scale bioacoustic geolocation.
    We formalize the problem, benchmark image-based and soundscape mapping methods on iNatSounds~\cite{chasmai2024inaturalist}, and propose a novel approach that uses species range predictions to improve geolocation.
    \item We investigate the potential of species information by constructing species oracles based on range estimation models. To study more realistic settings, we also introduce XCDC, a new dataset of species-rich dawn chorus recordings curated from \citet{xeno}.
    \item We demonstrate that spatiotemporal aggregation effectively mitigates key limitations of short recordings.
    \item We introduce multimodal geo-forensics through case studies of movie audio and imagery, illustrating potential real-world applications of multimodal geolocation.
\end{enumerate}

\section{Related Work}
\label{sec:related}

\shortparagraph{Audio Geolocation.}
The task of explicit geolocation of audio recordings remains largely underexplored. 
Several early studies~\cite{pokorny2019sound, kumar2016audio, choi2015multimodal, friedland2010multimodal, friedland2011multimodal} investigate audio and multimodal geolocation, but are limited to distinguishing sounds within or among specific regions. Motivated by the growing interest in image geolocation, we revisit and scale up the problem of audio geolocation from a modern perspective, aiming to address it at a global scale. 
Insights gained from our explorations in natural domains could also contribute to broader advancements in audio geolocation, as wildlife sounds such as birds and insects are also prevalent in urban settings.

\shortparagraph{Image Geolocation.}
Initial methods for image geolocation approached the problem through the lens of retrieval. To geolocate a ground image, they retrieved similar geo-tagged ground images~\cite{hays2008im2gps} or cross-view satellite images~\cite{workman2015wide, liu2019lending, shi2020looking, zhu2021vigor, yang2021cross, zhu2022transgeo}. 
Focus gradually shifted to feed-forward classification models that could directly predict location from an image by splitting the surface of the earth into several geo-cells and predicting the correct geo-cell~\cite{berton2022rethinking, weyand2016planet, vo2017revisiting, muller2018geolocation, clark2023we, pramanick2022world, seo2018cplanet}.
To overcome grid resolution constraints, some works explore adaptive grids~\cite{seo2018cplanet} and hierarchical approaches~\cite{clark2023we}. 
\citet{haas2024pigeon} incorporate political and administrative boundaries while creating their grids and beat one of the world’s foremost
professional GeoGuessr players~\cite{geoguessr}.

Following the release of CLIP~\cite{radford2021learning}, recent methods~\cite{klemmer2023satclip, vivanco2024geoclip} have revisited the retrieval approach.
These methods treat the location itself as a modality and learn location embeddings aligned with images.
GeoCLIP~\cite{vivanco2024geoclip} learns a location encoder that utilizes random Fourier features to capture high frequency details and learn better location features. 
SatCLIP~\cite{klemmer2023satclip} encodes location via spherical harmonics and learns a joint embedding space with paired satellite images. 
\citet{zhou2024img2loc} propose to leverage LLMs~\cite{achiam2023gpt, touvron2023llama} and incorporate domain knowledge with Retrieval Augmented Generation (RAG)~\cite{cai2022recent}.
In this work, we shift the focus from  visual to the acoustic modality. Audio presents a unique challenge, as it typically carries less explicit and globally consistent geographic information than visual data. We extend and benchmark multiple image geolocation methods across popular paradigms in an effort to establish a comprehensive benchmark for the novel task of bioacoustic geolocation.

\shortparagraph{Soundscape Mapping and Audio-Location Alignment.}
In the acoustic space, researchers are also interested in the complementary problem of soundscape mapping. Instead of identifying geographically unique signals in audio recordings, the focus is to learn common patterns in sounds from a particular location. 
The task is posed as a retrieval of audio conditioned on location. 
Prior research focus on understanding the soundscapes of either a few cities~\cite{aiello2016chatty} or for the entire earth~\cite{salem2018soundscape}. GeoCLAP~\cite{khanal2023learning} and PSM~\cite{khanal2024psm} use satellite images to represent geographic location and learn a shared embedding space between overhead images and ground audio. 
The domain of wildlife sounds also provides a unique set of challenges not as prominent in urban settings. 
Taxabind~\cite{sastry2025taxabind} learns a shared feature space for 6 modalities related to species and demonstrate its benefit to ecological problems. However, they too do not focus on audio geolocation and have \emph{implicit} audio-location pairing with ground images as an intermediary. 
While these methods can be implicitly used for 
geolocation, the unique challenges of global audio geolocation remain largely underexplored, a critical gap this work aims to fill.

\shortparagraph{Related Audio Tasks.}
In acoustic analysis, prior work also focus on localizing sources within specific scenes relative to the microphone position~\cite{sun2017indoor, 8794231, wu2021sslide, chung2022sound, zhang2018new, dang2019indoor, perotin2019regression, chen2020soundspaces, chen2021semantic, chen2023sound}.
While the structure of this problem is very similar to our study of audio geolocation, the two tasks require very different signals: features learned for one may not be predictive of the other.
There has also been some prior work in geolocating speech, either as a proxy task for identifying accents and dialects~\cite{van2016sprekend, lohfink2017sprekend, dehak2010front, foley2024you}, or to improve speech recognition (ASR)~\cite{xiao2018geographic, bell2015system}.
Our work focuses on the location of the location of the recording environment instead of the origin of a subject, for which human speech often provides insufficient or ambiguous geographic information.

\shortparagraph{Species Identification.}
The problem of identifying species from sounds has received significant attention in recent years, with approaches exploring both supervised~\cite{van2025perch, moummad2026domain} and self-supervised~\cite{wuao2026masked, rauch2025can} learning. 
~\citet{jeantet2023improving} assume access to location at test time, and learn to predict species conditioned on the location.
~\citet{chasmai2026metaperch} include a location  prediction task, but treat it only as an auxiliary training objective to aid species identification.
We focus on geolocation as a core problem and demonstrate the utility of species sounds in addressing it. 
In bioacoustics, species identification and geolocation are inherently coupled, and developments in either problem will help advance the other.

\begin{figure*}[t]
\begin{flushright}
\setlength{\tabcolsep}{0.117\linewidth}
\tiny
  \includegraphics[width=\linewidth]{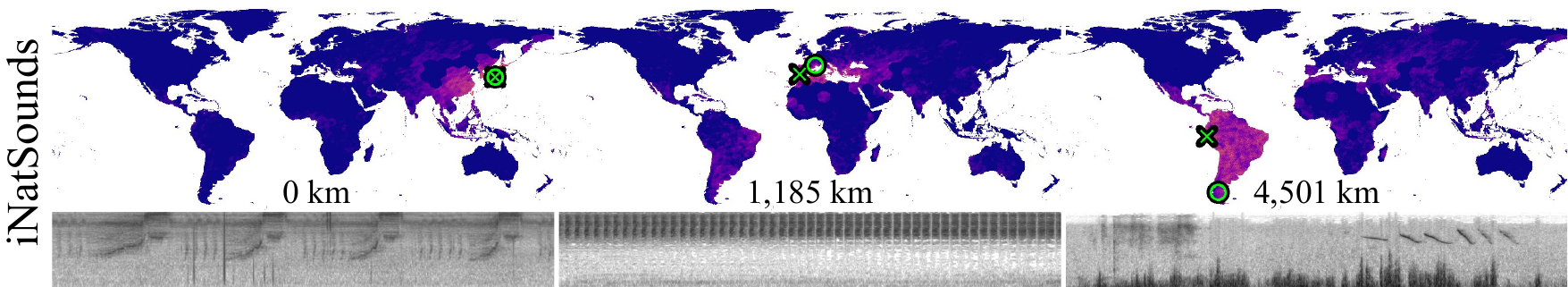}
  \begin{tabular}{ccc}
       \href{https://www.inaturalist.org/observations/182912774}{Observation URL} & 
       \href{https://www.inaturalist.org/observations/169912837}{Observation URL} & 
       \href{https://www.inaturalist.org/observations/192802995}{Observation URL} 
  \end{tabular}
  \includegraphics[width=\linewidth]{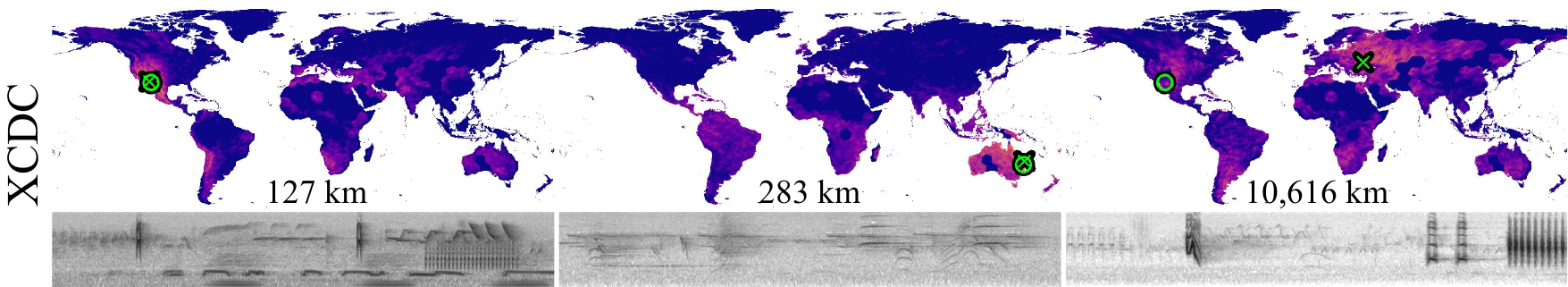}
  \begin{tabular}{ccc}
       \href{https://xeno-canto.org/268217}{Observation URL} & 
       \href{https://xeno-canto.org/505753}{Observation URL} & 
       \href{https://xeno-canto.org/267232}{Observation URL} 
  \end{tabular}
  \caption{
  \textbf{Geolocation Predictions.} Sample predictions of our method on iNatSounds~\cite{chasmai2024inaturalist} and XCDC. Heatmap shows the unnormalised likelihood of each location. Green crosses denote the final prediction (argmax) and green circles denote true location. 
  Geolocation error reported below each heatmap. 
  URLs to the original observations (clickable) provided below each spectrogram.
  Harshness of the region threshold (200km) can be better appreciated through the bottom left and center exemplars.
  We present additional visualizations in supplemental Fig.~\ref{fig:predictions_supp_inat}, \ref{fig:predictions_supp_xcdc}. 
  }
  \label{fig:predictions}
\end{flushright}
\end{figure*}

\section{Problem Setup and Methodology}
\label{sec:method}

In this work, we conduct experiments using iNatSounds~\cite{chasmai2024inaturalist}, a dataset of 230K audio recordings, with an average duration of around 20s.   
Each recording is annotated with a species, recording date, and geographic location. The training split includes audio from over 5,500 species and spans diverse regions across the Americas, Africa, Europe, Asia, and Australia. However, the distribution is imbalanced, with a bias toward major population centers. See supplementary Fig.~\ref{fig:class_grids} for the geographic distribution of iNatSounds and Fig.~\ref{fig:predictions} for some examples from around the world.

\shortparagraph{Preliminaries.}
Given an audio recording $\mathbf{x}$, our goal is to predict the coordinates $\mathbf{y}=(\text{latitude}, \text{longitude})$ of the location where $\mathbf{x}$ was recorded. 
We evaluate the geolocation error by computing the Haversine distance~\cite{gade2010non} between the predicted and true coordinates. 
Given a dataset ${\cal D} = \{(\mathbf{x}_i, \mathbf{y}_i)\}_{i=1}^N$  of audio recordings~$\mathbf{x}_i \in \mathcal{X}$ and ground truth locations $\mathbf{y}_i$, we measure the performance of a method using the median geolocation error across $\mathcal{D}$.
Following standard practice in image
geolocation~\cite{weyand2016planet, vivanco2024geoclip, haas2024pigeon},
we also report ``Percentage at Threshold,'' where a prediction is considered correct if the geolocation error is within a specific threshold. 
We use thresholds of 25 km, 200 km, 750 km, and 2,500 km, corresponding to city, regional, country, and continental scales.   
Refer to Fig.~\ref{fig:predictions} and supplementary Fig.~\ref{fig:scales} for illustrations of these scales.

We aim to train a geolocation model $g: \mathcal{X} \to \mathscr{L}$
where $\mathscr{L}$ is the space of all possible locations. 
This model is decomposed as $g = h_{\mathbf{\phi}} \circ f_{\mathbf{\theta}}$, where $f_{\mathbf{\theta}}: \mathcal{X} \rightarrow \mathbb{R}^d$, is an audio encoder with parameters $\mathbf{\theta}$, and $h_{\mathbf{\phi}}: \mathbb{R}^d \rightarrow \mathscr{L}$ is a decoder with parameters $\mathbf{\phi}$ responsible for geolocating an audio feature. 
The location can be encoded in different ways, and the definition of $\mathscr{L}$ changes accordingly. It can be simply (lat, lon) (for regression), a geographic bin label (for classification), or a learned embedding (for retrieval). 
Although the location encoding does not need to be reversible, each prediction in $\mathscr{L}$ should be mappable to a corresponding (lat, lon) for evaluation.

\shortparagraph{Audio Encoder.}
We adopt a vision-inspired approach to acoustic analysis by converting audio to a spectrogram ``image'' and extracting features with computer vision backbones. 
Model architecture and pre-training strategies are described in Sec.~\ref{sec:implementation} and spectrogram settings in supplemental Sec.~\ref{sec:supp:implement}. 
We ablate the audio encoders in Sec.~\ref{sec:main_ablations}. 

\shortparagraph{Retrieval for Geolocation.} 
We pose the problem of audio geolocation as retrieval, where $h_{\mathbf{\phi}}$ is responsible for computing the similarity of $f_{\mathbf{\theta}}(\mathbf{x})$ (query) with a predefined collection of embeddings (keys).
Recent prior work~\cite{vivanco2024geoclip} has explored location encoders $l_{\mathbf{\psi}} : \mathbb{R}^2 \rightarrow \mathbb{R}^d$, that learn dense representations for two-dimensional (lat, lon) values. We train $h_{\mathbf{\phi}}$ to project audio features into the embedding space of a location encoder. Retrieval can then be done at test time by comparing the predicted embedding $g(\mathbf{x})\in \mathscr{L}$ against a gallery of precomputed location embeddings, and returning the coordinates of the most similar location embedding. We use the training set itself to construct our gallery, so the predicted coordinates $\hat{\mathbf{y}}$ can be expressed as $\hat{\mathbf{y}} = \argmax_{\mathbf{y}_j \in \mathcal{D}^{Train}} g(\mathbf{x})^\top l_{\mathbf{\psi}}(\mathbf{y}_j)$.

\shortparagraph{Species Oracles.}\label{sec:checklist}
We briefly diverge to introduce a diagnostic tool designed to probe the limits of bioacoustic geolocation. We hypothesize that presence and absence of species is a strong signal for geolocation. 
One immediate challenge for testing this hypothesis is the unrealistic expectation of capturing sounds from all species in a location. 
In theory, if a recordist were to continuously capture audio at a location over an entire year, we might expect to record most species that occur there. 
With this complete ``checklist'' of species, how accurately can we predict recording location?

While year-long recordings are impractical, we can simulate these comprehensive checklists using precomputed species geographic range maps. 
For a given location on Earth, we determine which species' ranges contain that location and construct a checklist of species that are present or absent. 
Although expert-verified range maps are not available for all species, recent advancements~\cite{cole2023spatial, hamilton2024combining} have enabled joint prediction of range maps for tens of thousands of species. 
We leverage one such model (SINR~\cite{cole2023spatial}), to construct range maps for all species in the iNatSounds dataset. 
Finally, we learn $h_\phi$ to perform retrieval style geolocation using these checklists instead of audio features. Note that checklist construction assumes oracle access to the true location. These experiments do not reflect a realistic bioacoustic geolocation setting, but instead serve as upper bounds for species-based approaches.

\shortparagraph{AG-CLIP.} Our proposed method, \underline{A}udio\underline{G}eo-\underline{CLIP}, builds on top of GeoCLIP~\cite{vivanco2024geoclip} and includes an auxiliary task of predicting SINR~\cite{cole2023spatial} checklists from audio. Concretely, we add a checklist decoder and a BCE loss with the species checklists in addition to the usual location decoder $h_\phi$ and its contrastive loss. While a particular recording will likely capture only a small fraction of species whose ranges overlap the recording location, predicting the species checklist can encourage the encoder to attend not only to foreground vocalizations but also to background ambient sounds, learning to associate them with species likely to be encountered in that habitat. See supplemental Fig.~\ref{fig:supp_method} for a schematic. 

\shortparagraph{Handling Variable Length.}

For each recording, we construct a collection of 3s ``clips'' using a sliding window. Each clip is geolocated independently and then the average of all clip-predictions is used as the final prediction for that recording. We explore alternate poolings in Sec.~\ref{sec:main_ablations}.

\begin{table*}[t]
\setlength{\fboxsep}{0pt}
    \centering
    \caption{
    \small
    \textbf{Geolocation on iNatSounds}. 
    Experiments on iNatSounds test set. 
    The \colorbox{tablelightgray}{naive random predictor} and \colorbox{tablelightgray}{species oracles} contextualize performance. We train our models to geolocate using regression, classification, explicit sampling from range maps and retrieval. 
    Geolocation performance is evaluated by median error (km) and accuracies (\%) at different distance thresholds.
    We report mean $\textcolor{stdgray}{\pm\ std}$ from 3 runs. \colorbox{tabledarkgray}{\textdagger:  Off the shelf models}. 
    }
    \label{tab:main_neurips}
    \setlength{\tabcolsep}{9pt}
        \centering
        \begin{tabular}{clrcccc}
            \toprule
           \multicolumn{2}{c}{\multirow{2}{*}{\textbf{Experiment}}}  & \multicolumn{1}{c}{\textbf{$\downarrow$ Median}} & \textbf{$\uparrow$ City} & \textbf{Region} & \textbf{Country} & \textbf{Continent}\\
             &   & \multicolumn{1}{c}{\textbf{Error (km)}} & 25km & 200km & 750km & 2500km \\
             \midrule
             \rowcolor{tablelightgray} Naive & Rnd Train Loc & 7475 $\textcolor{stdgray}{\pm\ 32}$ & 00.1 $\textcolor{stdgray}{\pm\ 0.0}$ & 01.1 $\textcolor{stdgray}{\pm\ 0.0}$ & 06.6 $\textcolor{stdgray}{\pm\ 0.1}$ & 22.8 $\textcolor{stdgray}{\pm\ 0.3}$\\
            \midrule
             \rowcolor{tablelightgray} & Full checklist & 15 $\textcolor{stdgray}{\pm\ 00}$ & 64.0 $\textcolor{stdgray}{\pm\ 0.4}$ & 97.8 $\textcolor{stdgray}{\pm\ 0.0}$ & 99.9 $\textcolor{stdgray}{\pm\ 0.0}$ & 100. $\textcolor{stdgray}{\pm\ 0.0}$ \\
           \rowcolor{tablelightgray} & 50\% Corrupted & 54 $\textcolor{stdgray}{\pm\ 00}$ & 32.8 $\textcolor{stdgray}{\pm\ 0.0}$ & 83.3 $\textcolor{stdgray}{\pm\ 0.1}$ & 98.8 $\textcolor{stdgray}{\pm\ 0.0}$ & 100. $\textcolor{stdgray}{\pm\ 0.0}$\\
             \rowcolor{tablelightgray}\multirow{-3}{*}{\shortstack[c]{Species\\Oracles}} & Keep random 10 & 325 $\textcolor{stdgray}{\pm\ 02}$ & 10.0 $\textcolor{stdgray}{\pm\ 0.1}$ & 37.6 $\textcolor{stdgray}{\pm\ 0.2}$ & 76.6 $\textcolor{stdgray}{\pm\ 0.2}$ & 98.3 $\textcolor{stdgray}{\pm\ 0.0}$\\
            \midrule
             \multirow{2}{*}{Regress} & Euclidean & 1884 $\textcolor{stdgray}{\pm\ 05}$ & 00.1 $\textcolor{stdgray}{\pm\ 0.0}$ & 02.9 $\textcolor{stdgray}{\pm\ 0.0}$ & 23.4 $\textcolor{stdgray}{\pm\ 0.2}$ & 57.9 $\textcolor{stdgray}{\pm\ 0.1}$ \\
            & Haversine & 1602 $\textcolor{stdgray}{\pm\ 13}$ & 00.1 $\textcolor{stdgray}{\pm\ 0.0}$ & 04.2 $\textcolor{stdgray}{\pm\ 0.2}$ & 27.7 $\textcolor{stdgray}{\pm\ 0.3}$ & 61.8 $\textcolor{stdgray}{\pm\ 0.2}$ \\
            \midrule
            \multirow{3}{*}{Classify} & Res-0 (430 $\times 10^4$km$^2$)  & 1323 $\textcolor{stdgray}{\pm\ 09}$ & 00.0 $\textcolor{stdgray}{\pm\ 0.0}$ & 01.2 $\textcolor{stdgray}{\pm\ 0.0}$ & 22.6 $\textcolor{stdgray}{\pm\ 0.0}$ & 68.7 $\textcolor{stdgray}{\pm\ 0.2}$\\
            & Res-2 (8.6 $\times 10^4$km$^2$)  & 1326 $\textcolor{stdgray}{\pm\ 12}$ & 00.3 $\textcolor{stdgray}{\pm\ 0.0}$ & 16.3 $\textcolor{stdgray}{\pm\ 0.2}$ & 36.5 $\textcolor{stdgray}{\pm\ 0.4}$ & 64.1 $\textcolor{stdgray}{\pm\ 0.2}$\\
            & Hierarchical (0$\rightarrow$1$\rightarrow$2) & 1117 $\textcolor{stdgray}{\pm\ 06}$ & 00.3 $\textcolor{stdgray}{\pm\ 0.0}$ & 16.6 $\textcolor{stdgray}{\pm\ 0.2}$ & 40.0 $\textcolor{stdgray}{\pm\ 0.1}$ & 68.7 $\textcolor{stdgray}{\pm\ 0.2}$\\
            \midrule
            \multirow{3}{*}{\shortstack[c]{Species \\Ranges}} & Annotated Species & 1263 $\textcolor{stdgray}{\pm\ 01}$ & 00.3 $\textcolor{stdgray}{\pm\ 0.0}$ & 10.8 $\textcolor{stdgray}{\pm\ 0.1}$ & 35.7 $\textcolor{stdgray}{\pm\ 0.1}$ & 72.3 $\textcolor{stdgray}{\pm\ 0.1}$\\
            & Predicted Sp (Top 1) & 1664 $\textcolor{stdgray}{\pm\ 03}$ & 00.2 $\textcolor{stdgray}{\pm\ 0.0}$ & 08.2 $\textcolor{stdgray}{\pm\ 0.1}$ & 28.7 $\textcolor{stdgray}{\pm\ 0.0}$ & 62.5 $\textcolor{stdgray}{\pm\ 0.1}$ \\
            & Predicted Sp (All) & 1113 $\textcolor{stdgray}{\pm\ 02}$ & 00.3 $\textcolor{stdgray}{\pm\ 0.0}$ & 09.3 $\textcolor{stdgray}{\pm\ 0.1}$ & 37.7 $\textcolor{stdgray}{\pm\ 0.1}$ & \textbf{72.7} $\textcolor{stdgray}{\pm\ 0.1}$\\
            \midrule
            \rowcolor{tabledarkgray} \cellcolor{white}   & GeoCLAP\textsuperscript{\textdagger}  & 6856 $\textcolor{stdgray}{\pm\ 00}$ & 00.2 $\textcolor{stdgray}{\pm\ 0.0}$ & 01.3 $\textcolor{stdgray}{\pm\ 0.0}$ & 07.0 $\textcolor{stdgray}{\pm\ 0.0}$ & 24.9 $\textcolor{stdgray}{\pm\ 0.0}$\\
            \rowcolor{tabledarkgray} \cellcolor{white} & Taxabind\textsuperscript{\textdagger} & 4944 $\textcolor{stdgray}{\pm\ 00}$  & 00.4 $\textcolor{stdgray}{\pm\ 0.0}$ & 02.2 $\textcolor{stdgray}{\pm\ 0.0}$ & 11.9 $\textcolor{stdgray}{\pm\ 0.0}$ & 35.3 $\textcolor{stdgray}{\pm\ 0.0}$ \\
            \multirow{-3}{*}{Retrieve} & AG-CLIP (\textbf{ours})  & \textbf{1082} $\textcolor{stdgray}{\pm\ 11}$ & \textbf{06.4} $\textcolor{stdgray}{\pm\ 0.1}$ & \textbf{17.2} $\textcolor{stdgray}{\pm\ 0.1}$ & \textbf{41.0} $\textcolor{stdgray}{\pm\ 0.3}$ & 71.2 $\textcolor{stdgray}{\pm\ 0.2}$\\
            \bottomrule
        \end{tabular}%
\end{table*}

\section{Experiments}
\label{sec:experiments}

\subsection{Implementation Details}\label{sec:implementation} 
In the following experiments, the audio encoder $f_{\mathbf{\theta}}$ is a trainable MobileNet-V3~\cite{howard2019searching} model pretrained for species classification on the iNatSounds dataset. These models expect spectrograms that have been resized to 224x224x3. 
For \method, $h_{\mathbf{\phi}}$ is composed of two 2-layer MLPs, each with hidden dimensions of 128, responsible for projecting features to GeoCLIP embeddings and SINR checklists, respectively. In Sec.~\ref{sec:main_ablations}, we further explore the impact of different network architectures and pretraining strategies for both audio and location encodings. 
All models are trained using SGD with Nesterov acceleration. We use the validation set of iNatSounds for model selection. See Supp. Sec.~\ref{sec:supp:implement} for detailed model configurations.

\subsection{
Performance on iNatSounds
} \label{sec:inat_experiments}

We present the performance of geolocation methods on the iNatSounds test set in Table~\ref{tab:main_neurips}.

\shortparagraph{Naive Baseline.}
We start with a naive baseline that samples a random location from the training distribution for each test recording. This baseline yields a regional accuracy of just $1.1\%$, suggesting that the dataset is sufficiently diverse and not dominated by a few geographic hotspots.

\shortparagraph{Species Oracles.}
To contextualize subsequent results, we first present upper-bound performance using species checklists constructed with oracle access to the true location.
We binarize SINR~\cite{cole2023spatial} model outputs using a threshold of 0.1 to create binary vectors, or ``checklists'', indicating species presence and absence at each location. 
We then train a linear model to serve as $h_\phi$,
predicting GeoCLIP embeddings and geolocating via retrieval. 
Our results indicate that with a \emph{full checklist}, almost perfect region level performance is possible, confirming our hypothesis that species information serves as a strong geolocation signal. 
However, knowing all species that may be found at a location is not realistic with reasonably sized audio recordings and noisy or incomplete species identification. 
To simulate imperfect recall, we omit a random subset of species from the checklist. With \emph{50\% corruption}, regional accuracy remains relatively high at 83.3\%. If only \emph{10 randomly} selected species are retained, region performance drops to a still strong 37.6\%.

\shortparagraph{Regress.}
Regression methods tend to do the worst overall, with region level performances of 2.9\% and 4.2\% for Euclidean and Haversine, respectively. Haversine distance more accurately captures geolocation error, which is the likely cause of its better performance as a loss function.

\begin{figure*}[t]
  \centering
  \hfill
  \includegraphics[width=0.415\linewidth]{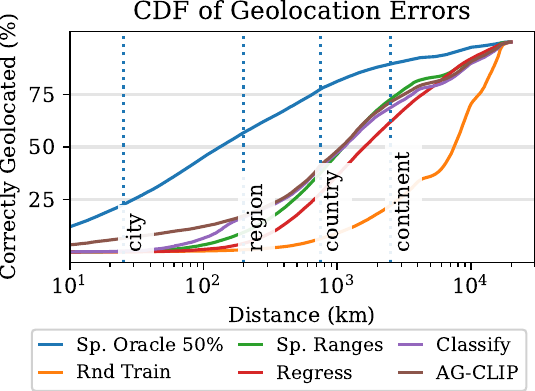}
  \hfill
  \includegraphics[width=0.57\linewidth]{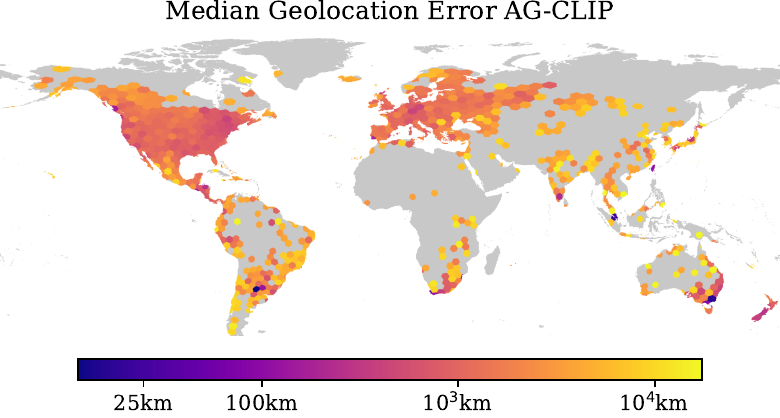}
  \hfill
  \caption{
  \small
  \textbf{Geolocation Error Trends.} 
  Left: Cumulative distribution of geolocation errors for \method and baseline models. Table~\ref{tab:main_neurips} accuracy metrics correspond to points on these curves. Right: Median of AG-CLIP geolocation errors binned by H3~\cite{brodsky2018h3} grid cells, capturing spatial variation in performance. 
  }
  \label{fig:plots}
\end{figure*}
\begin{table*}
\centering
\setlength{\tabcolsep}{6pt}
    \begin{subtable}[t]{0.39\textwidth}
    \begin{tabular}[t]{lcccc}
        \toprule
         Model & Pre & FT & Reg. & Cont. \\
        \midrule
           Wav2CLIP & VGG & \xmark & 03.8 & 30.3\\
           CLAP & LAION & \xmark & 05.1 & 34.6\\
           MobV3 & ImgNet & \checkmark & 14.4 & 64.9\\
           \rowcolor{tablegray}
           MobV3 & iNat & \checkmark & \textbf{17.2} & \textbf{71.2}\\
        \bottomrule
    \end{tabular}
    \caption{Audio Encoders}
    \label{tab:ablations:audio}
    \end{subtable}
    \hspace{\fill}
    \begin{subtable}[t]{0.32\textwidth}
    \begin{tabular}[t]{lccc}
        \toprule
         Experiment & City & Reg. & Cont. \\
        \midrule
            SatCLIP & 1.4 & 14.7 & 63.2\\
           SINR & 2.2 & 17.1 & 68.6\\
           GeoCLIP & \textbf{6.4} & 17.0 & 68.5\\
           \rowcolor{tablegray}" + checklist  & \textbf{6.4} & \textbf{17.2} & \textbf{71.2} \\
        \bottomrule
    \end{tabular}
    \caption{Location Encoders}
    \label{tab:ablations:location}
    \end{subtable}
    \hspace{\fill}
    \begin{subtable}[t]{0.26\textwidth}
    \begin{tabular}[t]{lcc}
        \toprule
         Experiment & Reg. & Cont. \\
        \midrule
           \rowcolor{tablegray}Average & \textbf{17.2} & 71.2\\
           Max Pool & 16.5 & 68.6\\
           Cluster & 15.5 & 67.9 \\
           Transformer & 16.8 & \textbf{71.5}\\
        \bottomrule
    \end{tabular}
    \caption{Variable Length}
    \label{tab:ablations:agg}
    \end{subtable}
\caption{\small\textbf{Ablations.} Importance and alternative choices of different components of AG-CLIP. We include Region (Reg. 200km) and Continent (Cont. 2500km) performance on iNatSounds test set. 
}
\label{tab:ablations}
\end{table*}

\shortparagraph{Classify.} For classification, we use the H3 library~\cite{brodsky2018h3} to divide the world into a hexagonal grid. 
$h_{\mathbf{\phi}}$ is implemented as an N-way classifier where N is the number of hexagons in the grid. 
We experiment with two grid resolutions, defined by the area of each cell. More details and visualization of these grids are presented in Supp. Sec.~\ref{sec:supp:classification}. 
Lower resolution (bigger cells) leads to better continental performance while higher resolution leads to better regional and country level performance. 
A hierarchical approach gets the best of both, performing 
similar to low resolution for continent (68.7\% vs 68.7\%) and better than high resolution for country (40.0\% vs 36.5\%).

\shortparagraph{Species Ranges.} We next evaluate performance when explicit species information is available. Each iNatSounds recording includes a species annotation. By sampling the species’ SINR~\cite{cole2023spatial} likelihood distribution over a geographic grid, we can geolocate the recording. Note that SINR is trained on a much larger corpus of species occurrence data, which may offer an advantage over other methods.
Using the annotated species, we see a region accuracy of $10.8\%$ and a continent accuracy of $72.3\%$. 
However, the availability of ground-truth target species annotations at test time is unrealistic; instead, we can train a classifier to predict the target species from the audio. 
Using the SINR distribution of the top-1 predicted target species drops the region and continent performance to $8.2\%$ and $62.5\%$ respectively.
Rather than using a single predicted species for each recording, we can use the per-species scores given by our species classifier to weigh and combine SINR likelihood maps for all species.  
This weighted combination of species likelihood maps is slightly better than even the \emph{annotated} target species (country $37.7\%$ vs $35.7\%$), 
which may be an artifact of these classifiers' ability to capture background species to some extent~\cite{chasmai2024inaturalist}.

\begin{table*}[t]
    \caption{
    \textbf{Species-Rich Audio with XCDC}.  Left: Models trained on iNatSounds training set and evaluated on XCDC. We report mean of 3 runs. 
    Right: Geolocation with species ranges for randomly sampled subsets of ground truth species. We plot mean, std and range (min-max) over 100 runs. 
    }
    \label{tab:xcdc}
    \setlength{\tabcolsep}{4pt}
    \hfill
    \begin{minipage}{0.58\textwidth}
        \begin{tabular}{lrcccc}
            \toprule
             \multirow{1}{*}{\textbf{Experiment}} & \multicolumn{1}{c}{\textbf{Error}} & \textbf{City} & \textbf{Region} & \textbf{Cou.} & \textbf{Cont.}\\
            \midrule
            Species Ranges (True) & 449 & 00.5  & 25.8  & 68.8  & 99.0  \\
             Species Ranges (Predicted) &  1097 & 00.1  & 02.5  & 27.6  & 80.0  \\
            \midrule
            Classification (Hierarchical) & 1116 & 00.0  & 05.3  & 31.7  & 69.7  \\
              \midrule
             AG-CLIP (\textbf{ours}) & 1112 & 00.2  & 04.3  & 26.3  & 71.9 \\
            \bottomrule
        \end{tabular}%
    \end{minipage}
    \hfill
    \begin{minipage}{0.36\textwidth}
        \includegraphics[width=\textwidth]{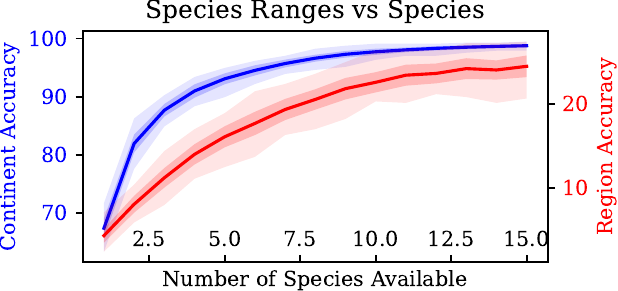}
    \end{minipage}
    \hfill
    
\end{table*}

\begin{table}[t]
    \centering
    \caption{
    \textbf{Spatiotemporal Aggregation}. Left: AG-CLIP geolocation can be improved by grouping recordings from the same neighborhood, over a year, month or week.  
    }
    \label{tab:space_agg}
    \setlength{\tabcolsep}{2.3pt}
    \centering
        \centering
        \begin{tabular}{lrcccc}
            \toprule
            \textbf{Experiment} &  \multicolumn{1}{c}{\textbf{Error}} & \textbf{City} & \textbf{Reg.} & \textbf{Cou.} & \textbf{Cont.}\\
            \midrule
            No Grouping & 1082 & 06.4 & 17.2 & 41.0 & 71.2 \\
            \midrule
            Res-6 (36 km$^2$) + Week & 890 & 07.9 & 19.9 & 45.8 & 76.5 \\
            Res-6 (36 km$^2$) + Month & 802 & 08.5 & 21.2 & 48.2 & 78.3 \\ 
            Res-6 (36 km$^2$) + Year & 651 & 10.8 & 25.9 & 55.6 & 84.0 \\
            Res-5 (253km$^2$) + Year & 520 & 13.2 & 30.4 & 62.1 & 88.2 \\
            \bottomrule
        \end{tabular}
        \vspace{-1mm}
\end{table}

\shortparagraph{Retrieve.} For retrieval, we explore methods that jointly learn audio and location encoders. 
GeoCLAP~\cite{khanal2023learning} and Taxabind~\cite{sastry2025taxabind} are soundscape mapping methods that have their own audio and location encoders, which we use off-the-shelf without any additional training. 
GeoCLAP performs poorly, likely due to domain shift, as it was trained on urban sounds from SoundingEarth~\cite{heidler2023self}. On the other hand, Taxabind, which was trained on a different subset of iNaturalist~\cite{iNaturalist}, achieves better results with a country level accuracy of 11.9\%.
However, it still underperforms relative to our models, likely because our models benefit from explicitly paired audio and location data, whereas Taxabind relies on only the implicit pairing available through images. 
Our method, AG-CLIP, achieves the best performance overall, with region and country level performances of 17.2\% and 41.0\%, respectively.

\shortparagraph{Performance at Different Distance Thresholds.}
Fig.~\ref{fig:plots} (left) shows CDF curves for various methods, plotting the fraction of test recordings (y-axis) correctly geolocated within a given distance (x-axis). 
These curves reveal how models perform across geospatial scales. 
Geolocation based on species checklists, even with 50\% corruption, is best across scales. 
Among audio geolocation methods, retrieval based AG-CLIP does best at finer scales like the city and region. Hierarchical classification shows a jump in performance around the region level and is close to AG-CLIP for coarser scales. This may be due to the finest grid cells being similar in size to a region, limiting the model’s ability to predict at finer resolutions. Explicit species identification with range analysis is relatively poor for finer scales, but overtakes AG-CLIP around the 1500\,km mark.

\shortparagraph{Spatial Variation.}
Fig.~\ref{fig:plots} (right) shows how geolocation performance of AG-CLIP varies geospatially. 
We associate each recording with a resolution 2 hexagon from H3~\cite{brodsky2018h3} and compute the median error per hexagon. Darker (bluer) regions indicate better geolocation. 
The model performs better in regions with more training data, like the US and Europe. Interestingly, certain areas such as Central America, South Africa, Taiwan, Eastern Australia, and New Zealand show particularly good performance, possibly hinting at the presence of distinctive soundscapes here.

\subsection{Ablations 
}
\label{sec:main_ablations}

We ablate the audio encoders, location encoders, and temporal aggregation strategies in Table~\ref{tab:ablations}. 
Off-the-shelf audio encoders like Wav2CLIP~\cite{wu2022wav2clip} and CLAP~\cite{elizalde2023clap} perform worse than pretraining on iNatSounds (Table~\ref{tab:ablations:audio}).
Starting with ImageNet pretrained weights instead of iNatSounds also reduces performance, with $7\%$ and $3\%$ drops at the continent and region levels. 
 
Next, we ablate the location encoder in Table~\ref{tab:ablations:location}. With SatCLIP~\cite{klemmer2023satclip}, we observe consistently worse performance than SINR~\cite{cole2023spatial} or GeoCLIP~\cite{vivanco2024geoclip}. SINR is comparable to GeoCLIP at the region and continent levels, but tapers off at the city level.
Better performance of GeoCLIP may be because of its alignment with ground-level imagery, unlike SatCLIP (satellite images) or SINR (species observations).
The auxiliary checklist loss introduced in AG-CLIP improves performance by 2.7\% at the continent level, though the gains diminish at finer scales. 

Finally, we explore alternate pooling methods to handle variable length audio in Table~\ref{tab:ablations:agg}. Our default is a simple average.
We see a drop in performance if max-pooling is used instead. We could also first cluster the embeddings via K Means (k=5) and then use the centroid of the largest cluster as the aggregate. While better than max-pooling, this is still worse than averaging. 
We also try training a transformer model to pool frozen embeddings (see supplemental Sec.~\ref{sec:supp:transformer} for more details). 
The transformer performs well at the continent level, but we prefer averaging because of its simplicity and lack of additional training.

\begin{figure*}[t]
\centering
  \includegraphics[width=\linewidth]{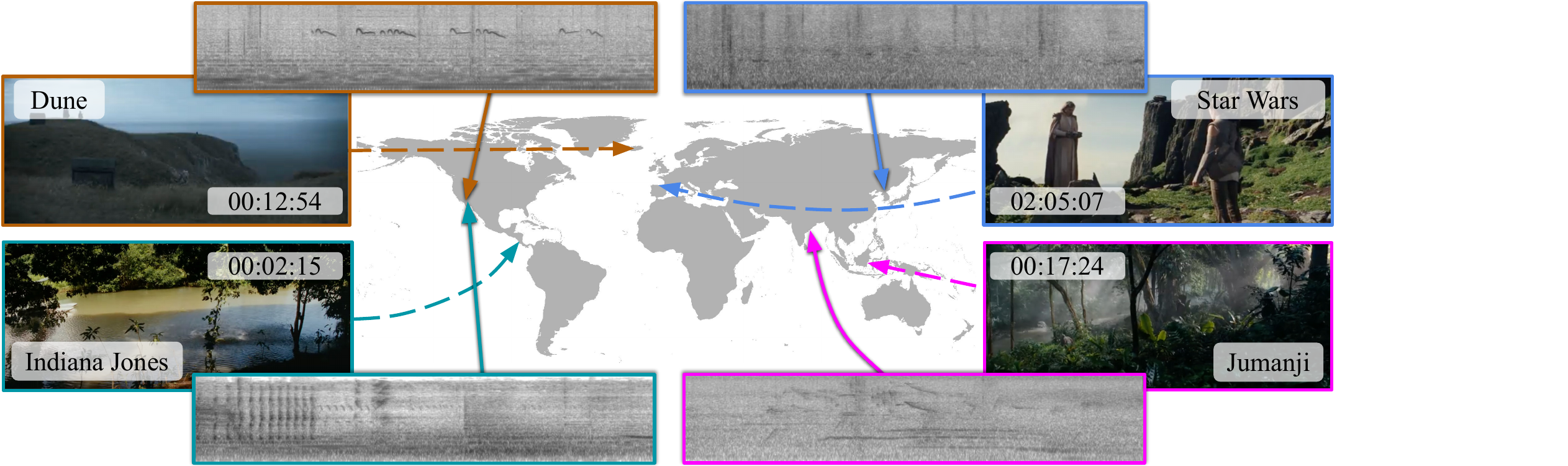}
  \caption{
  \textbf{Multimodal Geo-Forensics with Movie Clips.} 
  Geolocation predictions for audio and video frames from scenes in four movies. 
  Arrows on the map indicate the predicted locations for each modality. Discrepancies between modalities reveal potential artifacts introduced during post-production. 
  }
  \label{fig:forensics}
\end{figure*}

\subsection{XCDC and Spatiotemporal Aggregation}
\label{sec:xcdc_space}

Our species oracle experiments (Table~\ref{tab:main_neurips}) demonstrate that geolocation accuracy increases with more knowledge of the species present at the recording location. 
However, short recordings often capture only a few species, limiting this important signal. 
To evaluate whether increased species diversity improves geolocation, we explore two strategies: (1) sampling long, species-rich recordings from the dawn chorus and (2) aggregating short recordings from the same spatiotemporal neighborhood.

\shortparagraph{Species-Rich Audio.}
We construct a dataset of species-rich audio by sampling dawn chorus recordings from \citet{xeno}, a global archive of natural sound recordings. 
The dawn chorus is a period of intense, multi-species vocal activity that occurs near sunrise, particularly during the breeding season~\cite{gil2020bird, weldy2024audio}. 
To isolate these species-rich recordings, we select audio that (1) was recorded during the spring dawn chorus, (2) is at least 3 minutes in duration, and (3) contains annotations for at least 10 distinct species. 
This filtering yields 576 recordings with associated geographic coordinates and species labels. 
We refer to this as Xeno-Canto Dawn Chorus, a new benchmark for evaluating geolocation from species-rich soundscapes.

Our results are shown in Table~\ref{tab:xcdc}. 
We reconfirm the value of species information: geolocation accuracy increases as more ground truth species are provided (Table~\ref{tab:xcdc}, right), with full species knowledge yielding a region-level accuracy of 25.8\% (Table~\ref{tab:xcdc}, top row). 
However, model-based approaches that rely on predicted species ranges, hierarchical classification, or retrieval achieve substantially lower performance ($\le 5.3\%$), roughly equivalent to knowing only a single species. We observe similar trends for WABAD~\cite{perez2025wabad}, an in-the-wild, passive acoustic monitoring dataset (See supplemental Sec.~\ref{sec:supp:wabad}).
This suggests that current models struggle to exploit the multi-species signals present in these complex soundscapes. 
A likely explanation is the distribution shift: most iNatSounds recordings contain fewer species, and the models may not generalize well to audio with significant species overlap.

\shortparagraph{Spatiotemporal Aggregation.}
We group iNatSounds test recordings by location and time to simulate spatiotemporal aggregation. 
Spatial grouping is performed using H3 hexagons at resolutions 5 and 6, while temporal grouping spans the full year, individual months, or individual weeks. 
To geolocate a group, we first apply our model independently to each recording, then average the predicted location distributions across the group. 
The aggregated prediction is assigned to all recordings in the group, allowing direct comparison with recording-level results from Table~\ref{tab:main_neurips}.

Table~\ref{tab:space_agg} reports AG-CLIP performance under different spatiotemporal aggregation strategies. 
When aggregating recordings within $36\,\text{km}^2$ neighborhoods and a week, the region performance improves to $19.9\%$. This modest improvement reflects a limited aggregation: 
over 80\% of groups in the weekly grouping contain only a single species. 
When aggregation is extended to a month, performance improves to $21.2\%$. 
Aggregating recordings across the full year within $36\,\text{km}^2$ and $253\,\text{km}^2$ neighborhoods improves region-level accuracy from $17.2\%$ (no grouping) to $25.9\%$ and $30.4\%$, respectively. 
These gains relative to the ungrouped baseline suggest that the model is able to leverage complementary species information spread across multiple recordings.

Both XCDC and spatiotemporal aggregation involve recordings with more species information than iNatSounds. A key distinction is that while the soundscape recordings in the former would have significant species overlap and background noise, the latter would yield curated recordings with distinct vocalizations. The substantially worse performance on XCDC suggests that species overlap is a key factor limiting geolocation performance.

\subsection{Multimodal Geo-Forensics}
\label{sec:geo_forensics}
Live broadcasts of events, such as professional golf tournaments, often enhance their ambiance by inserting bird vocalizations. 
However, observers have noted cases where producers included species not found at the event location~\cite{golf}. 
Similar mismatches arise in film, where audio effects may not correspond to the geographic setting depicted on screen~\cite{dune, jones}. 
These inconsistencies offer a unique opportunity for \textit{multimodal geo-forensics}.

In Fig.~\ref{fig:forensics}, we analyze scenes from four Hollywood films that had been previously flagged by viewers for mismatches between visual setting and soundscape. 
In all cases, audio- and image-based geolocation produced divergent predictions, confirming the presence of modality inconsistencies. 
Jumanji showed the closest alignment between modalities at 4480 km, while Star Wars exhibited the largest divergence at 9092 km. 
These qualitative results highlight the potential of multimodal geo-forensics as a novel application domain, leveraging cross-modal cues to identify geographic inconsistencies. 
We view this as a promising direction for future work at the intersection of vision, sound, and place.
See supplemental Sec.~\ref{sec:supp:geo_forensics} for more~details.

\section{Discussion}

\paragraph{Environmental and Anthropogenic Sounds.}
Although iNatSounds is primarily constructed for species identification, it contains other unannotated sound sources including environmental sounds such as wind, rain, flowing water, and rustling leaves; as well as anthropogenic sounds such as human speech and vehicle horns. 
These sources offer complementary context to bioacoustic signals, and help refine the location estimate.

Our species range experiments (Tab~\ref{tab:main_neurips})
help disentangle the contributions of these signals to overall geolocation performance. The end-to-end (audio$\rightarrow$loc) model AG-CLIP generally performs better than the species-only model (audio$\rightarrow$species$\rightarrow$loc) and we suspect that this gap is primarily due to environmental and anthropogenic sounds. The performance gap is more pronounced at finer resolutions---AG-CLIP is better by 6\%, 8\% and 3\% at the city, region, and country levels respectively.

This disentanglement is also visible in our audio encoder ablations in Table~\ref{tab:ablations:audio}. Audio encoders such as Wav2CLIP and CLAP are trained on general audio datasets, and their embeddings ought to capture environmental and anthropogenic sounds to an extent, while missing fine-grained details of species sounds. The performance gap between these general audio models and our approach reflects the contribution of bioacoustic signals in our overall geolocation performance.

\shortparagraph{Geolocating Audio is Challenging.}
Although AG-CLIP outperforms the baselines, its absolute regional performance remains low at 17.2\%, which we attribute to several key factors.
First, species ranges can be extremely large, sometimes spanning entire continents. Recordings capturing a small number of species may not have enough information for precise geolocation. This can also be seen in our Species Ranges - Annotated experiment in Table~\ref{tab:main_neurips}---using the annotated focal species yields a low regional accuracy of 10.8\%.

Second, the world is acoustically self-similar across distant regions, and our method often makes errors which correspond to regions that have similar geographical context, but are far spatially. An example can be seen in the first XCDC case in Fig~\ref{fig:predictions}. The recording is from southwestern North America, and our model prediction is reasonably good. However, it also assigns high likelihoods to very distant regions in North Africa and West Asia exhibiting broadly similar arid climates.

Finally, we believe geolocating audio is inherently more challenging than images. Useful signals for image geolocation—such as famous landmarks and architectural style---are simply inaccessible because they do not make sounds. While certain geographically varying audio sources---such as sirens of emergency vehicles---are informative, short durations of audio recordings limit the likelihood of capturing them.  This inherent challenge is reflected in the gap between our region performance of 17.2\% and corresponding 31-63\% accuracies reported by recent approaches~\cite{haas2024pigeon} on different image geolocation benchmarks.

\section{Conclusion and Future Work}
\label{sec:conclusion}

Geolocating arbitrary audio is challenging, but recordings containing bioacoustic signals offer promising geographic cues.  
Our experiments demonstrate the strong potential of species-centric geolocation, and highlight the gap in performance of current approaches. 
Although short recordings may lack enough information for precise geolocation, we show that aggregating audio across space and time, a practical approach for applications like iNaturalist and Merlin, substantially improves performance. 
Our models transfer to real-world settings, but species overlap and difficulty in identification limit performance. Overall, the task remains substantially more challenging than image geolocation, leaving significant scope for improvement and future exploration.

Future work could improve species classification in complex soundscapes, develop architectures that better capture long-range acoustic context, and explore structured fusion across multiple recordings.  
Collecting training data with dense species sounds may help address distributional shifts relative to datasets like XCDC. Better sampling or reweighting strategies could help alleviate the effects of geographical biases in iNatSounds. Incorporating seasonality into geolocation models remains an important open direction.  
Our multimodal geo-forensics case studies highlight opportunities for combining audio and visual signals, opening new avenues for future exploration.

In the long-term, we envision bioacoustic geolocation enabling impactful applications, from providing geospatial context in applications such as Merlin Sound ID to habitat change monitoring via unlabeled soundscapes. 
Aggregating evidence across a long recording, or across multiple recordings from neighboring locations is a promising direction to overcome the limitation imposed by short audio recordings. Extension to general audio could be useful for applications like localizing distress calls and digital forensics.  More broadly, just as image geolocation opened the door to a wide range of use cases, we believe audio geolocation is an emerging capability that will find its place as tools and datasets mature. The scientific challenge alone is also compelling: can we localize a place based solely on what it sounds like?

\section*{Acknowledgements}
The project was funded in part by NSF Grant \#2329927.

\section*{Impact Statement}

\shortparagraph{Data Usage and Licensing.} The observations in iNatSounds are licensed for research use~\cite{chasmai2024inaturalist}. We do not modify or re-release any data from iNatSounds. 
For XCDC, we collect recordings from openly available data in XenoCanto. As per their website, the usage of each recording can be specified by the user who uploaded it. From the recordings in our filtering criteria, we pick the most restrictive one (CC BY-NC-ND), and release XCDC with this license. For the recordings that we present as prediction visualizations (Fig~\ref{fig:predictions},  ~\ref{fig:predictions_supp_inat} and ~\ref{fig:predictions_supp_xcdc}), we link the original observations and credit the users who contributed those recordings.

Human voice or other personally identifiable content can be present in iNatSounds and XCDC. Since the recordings are openly available and licensed for research use, we do not obfuscate or modify the data for anonymization. 

\shortparagraph{Privacy Concerns.} An adversary with the ability to infer recording locations of any online content just from the audio can violate the creator’s privacy. On the other hand, by understanding the geographic cues embedded in audio, it may be possible to develop methods to suppress or obfuscate them, thereby mitigating such risks. At its current stage, our work focuses on natural sounds, and even our best models can get the city correct only ~6\% of the time. This suggests that the immediate privacy risks are somewhat limited. That said, we recognize that as audio geolocation models improve, these risks will grow and become an important consideration for the field.

\shortparagraph{Geographical Biases.}   iNatSounds exhibits geographic bias towards Western countries, particularly in the Northern Hemisphere. Our models exhibit better geolocation performance in regions with greater training data coverage (Fig.~\ref{fig:plots}). As the evaluation split of iNatSounds exhibits similar biases, the reported performance may be overestimated. 

\shortparagraph{Reproducibility}
Upon acceptance, we will release all code and detailed instructions for benchmarking geolocation models as well as for training our method and baselines. iNatSounds~\cite{chasmai2024inaturalist} is already publicly available and we will release XCDC with appropriate licensing and links to original \citet{xeno} observations.

\bibliography{main}
\bibliographystyle{icml2026}

\newpage
\appendix
\onecolumn

\appendix
\setcounter{table}{0}
\renewcommand{\thetable}{A\arabic{table}}
\setcounter{figure}{0}
\renewcommand{\thefigure}{A\arabic{figure}}

\section{Appendix}

We start with a visualizatoin of the distance thresholds used by our evaluation metrics in Sec.~\ref{sec:supp:scales}. We follow this with additional prediction visualizations in Sec.~\ref{sec:supp:predictions}. Next, we visualize the different ways to represent location for classification (Sec.~\ref{sec:supp:classification}) and retrieval (Sec.~\ref{sec:supp:galleries}). We next describe additional implementation details left out in the main paper (see Sec.~\ref{sec:supp:implement}). In particular, we elaborate on the methodology with a block diagram, describe the spectrogram creation process, and report tuned hyperparameters and other configurations. We also share additional details for GeoCLAP and TaxaBind as well as the transformer pooling ablation (Sec.~\ref{sec:supp:transformer}) in Table 2 (main paper). Next, in Sec.~\ref{sec:supp:ablation}, we report additional ablation on Location Galleries.  Additional results and standard deviations for XCDC are reported and discussed in Sec.\ref{sec:supp:xcdc}. We also experiment with an in-the-wild passive acoustic monitoring dataset in Sec.~\ref{sec:supp:wabad}. Next, we report additional details for geoforensics experiments (Sec.~\ref{sec:supp:geo_forensics}) and present additional experiments on multimodal geolocation (Sec.~\ref{sec:supp:multimodal}), species affinities (Sec.~\ref{sec:supp:affinity}) and multimodal retrieval (Sec.~\ref{sec:supp:audio2image}). 

Next, we present a taxanomic breakdown of our performance (Sec.~\ref{sec:supp:taxon_split}), followed by additional ablations on other bioacoustic encoders (Sec.~\ref{sec:supp:bio_encoders}) and architectural variants of our encoder (Sec.~\ref{sec:supp:architectures}).
We also briefly explore potential future directions such as data resampling to alleviate the effect of some geographical biases in the training dataset (Sec.~\ref{sec:supp:data_resampling}), and inclusion of time as an additional modality to aid geolocation (Sec.~\ref{sec:supp:time_modality}).
Next, we include some additional analyses of our models: 1) we analyze the quality of embeddings learned by our models (Sec.~\ref{sec:supp:embeddings}) and 2) we explore the relation between uncertainty in SINR checklists and our geolocation performance (Sec.~\ref{sec:supp:sinr_uncertain}). 
We conclude with a discussion on compute resources in Sec.~\ref{sec:supp:compute}.

\begin{figure}[ht]
  \centering
  \includegraphics[width=\linewidth]{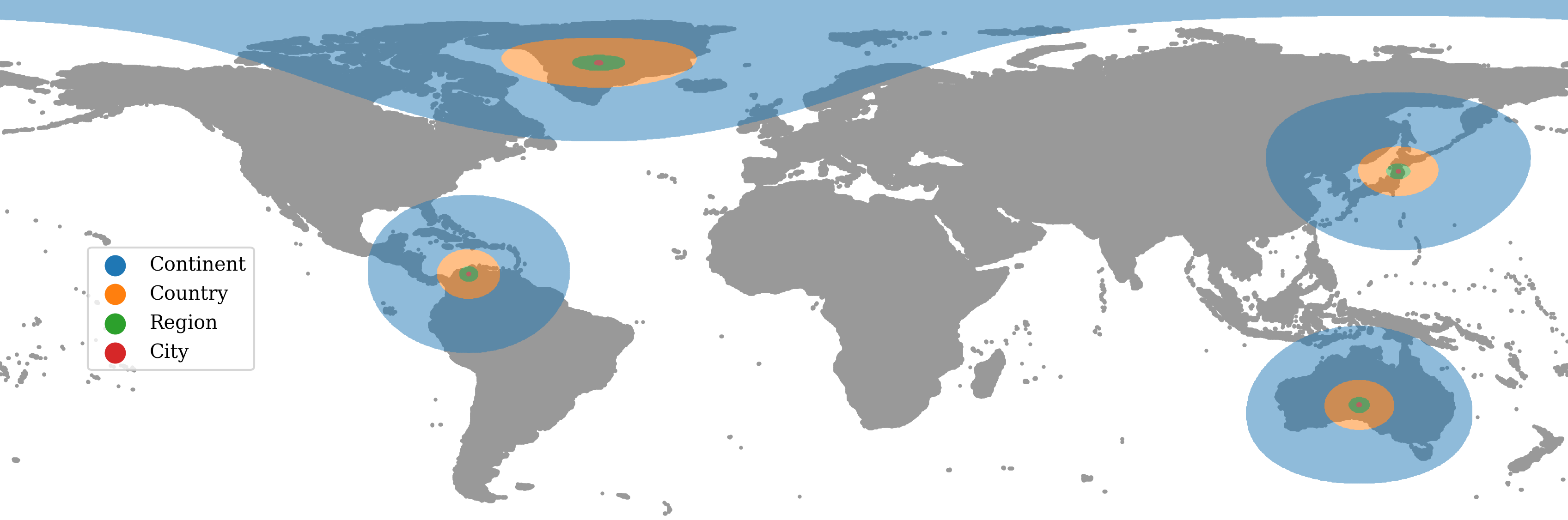}
  \caption{
  \textbf{Scales of Geolocation.}
  We plot all points within different thresholds used for calculating our geolocation metrics. These are not circles because we use haversine distance. We show these scales centered at a few different locations since the size changes at different points on earth. 
  }
  \label{fig:scales}
\end{figure}

\subsection{Scales of Geolocation}\label{sec:supp:scales}

For better understanding of the distance thresholds used in evaluation and an appreciation of the difficulty of this task, please see their visualization in Fig~\ref{fig:scales}. A prediction is considered correct at a given level if it falls within the corresponding area centered on the ground-truth location. 
Note the great difference in scale between finer levels like the city (barely bigger than a dot) and coarser levels like the country and continent.
While 200km for the region level may sound large, the small green areas in Fig~\ref{fig:scales} show how small it is compared to the size of the world. 
Even the continent threshold is actually harsher than it sounds, and covers only about the size of Australia.

\subsection{More Prediction Visualizations}\label{sec:supp:predictions}

\begin{figure}[t]
\centering
\setlength{\tabcolsep}{0.116\linewidth}
\tiny
    \includegraphics[width=\linewidth]{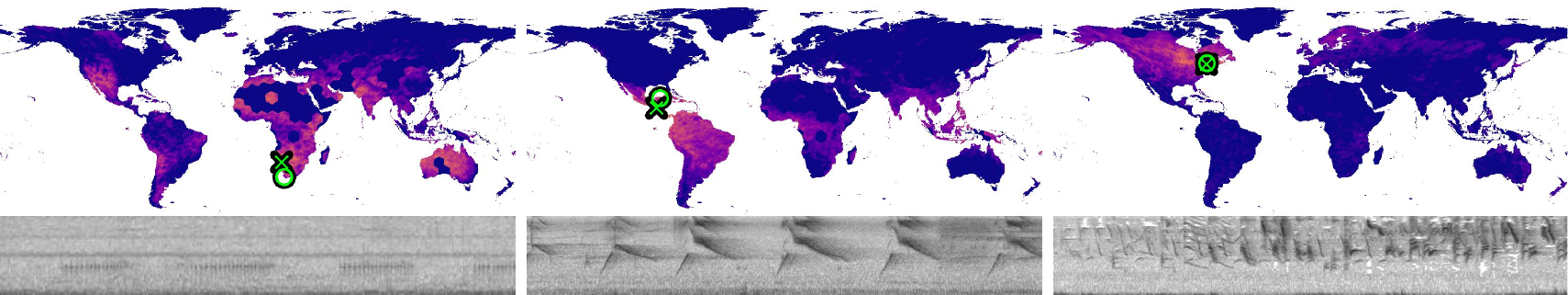}
    \begin{tabular}{ccc}
   \href{https://www.inaturalist.org/observations/146483231}{Observation link} & \href{https://www.inaturalist.org/observations/157249249}{Observation link} & \href{https://www.inaturalist.org/observations/161067490}{Observation link}
    \end{tabular}
    \includegraphics[width=\linewidth]{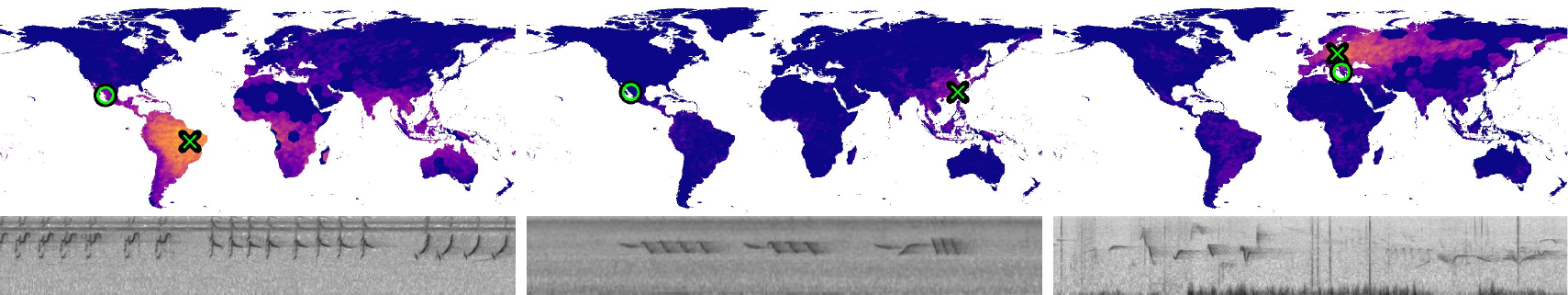}
    \begin{tabular}{ccc}
    \href{https://www.inaturalist.org/observations/159939962}{Observation link} & \href{https://www.inaturalist.org/observations/175865796}{Observation link} & \href{https://www.inaturalist.org/observations/155444921}{Observation link}
    \end{tabular}
    \includegraphics[width=\linewidth]{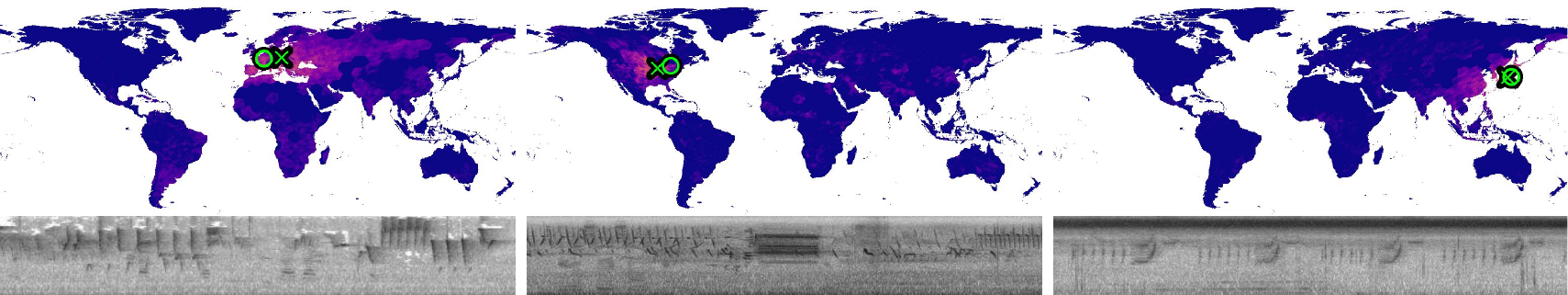}
    \begin{tabular}{ccc}
    \href{https://www.inaturalist.org/observations/163814982}{Observation link} & \href{https://www.inaturalist.org/observations/161525799}{Observation link} & \href{https://www.inaturalist.org/observations/175628923}{Observation link}
    \end{tabular}
    \caption{
  \textbf{Model Predictions on iNatSounds~\cite{chasmai2024inaturalist}.}  Heatmap shows the unnormalised likelihood of each location. Green crosses denote the final prediction (argmax) and green circles denote true location. Original observations linked below. 
  Best viewed zoomed in. 
  }\label{fig:predictions_supp_inat}
\end{figure}

\begin{figure}[t]
\centering
\setlength{\tabcolsep}{0.116\linewidth}
\tiny
    \includegraphics[width=\linewidth]{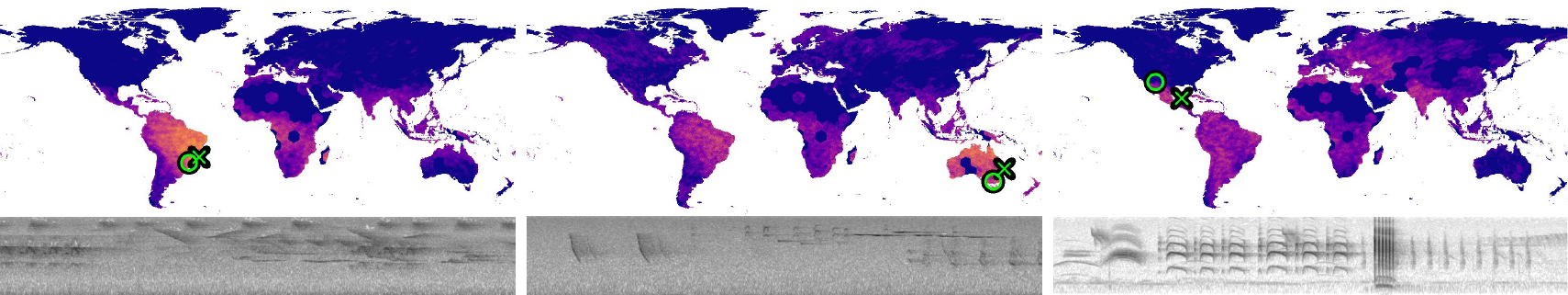}
    \begin{tabular}{ccc}
    \href{https://xeno-canto.org/624747}{Observation link} & \href{https://xeno-canto.org/641433}{Observation link} & \href{https://xeno-canto.org/254806}{Observation link}
    \end{tabular}
    \includegraphics[width=\linewidth]{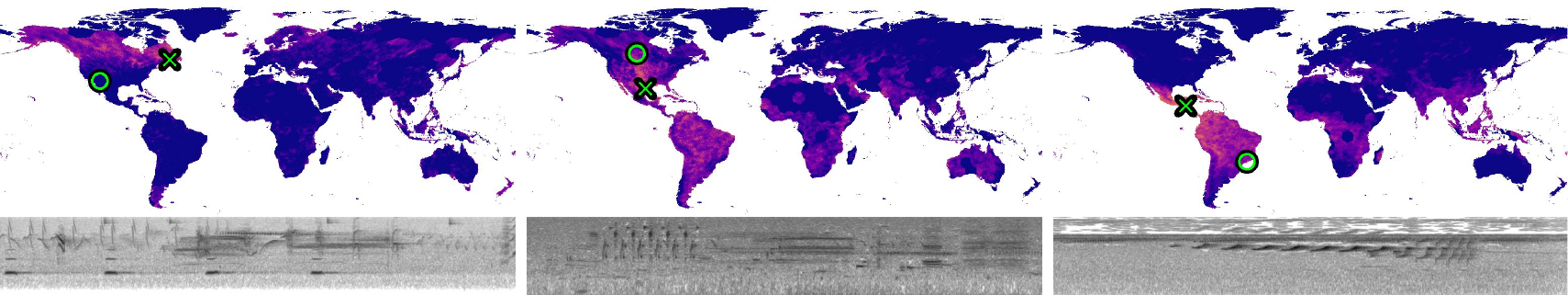}
    \begin{tabular}{ccc}
    \href{https://xeno-canto.org/322356}{Observation link} & \href{https://xeno-canto.org/153335}{Observation link} & \href{https://xeno-canto.org/391046}{Observation link}
    \end{tabular}
    \includegraphics[width=\linewidth]{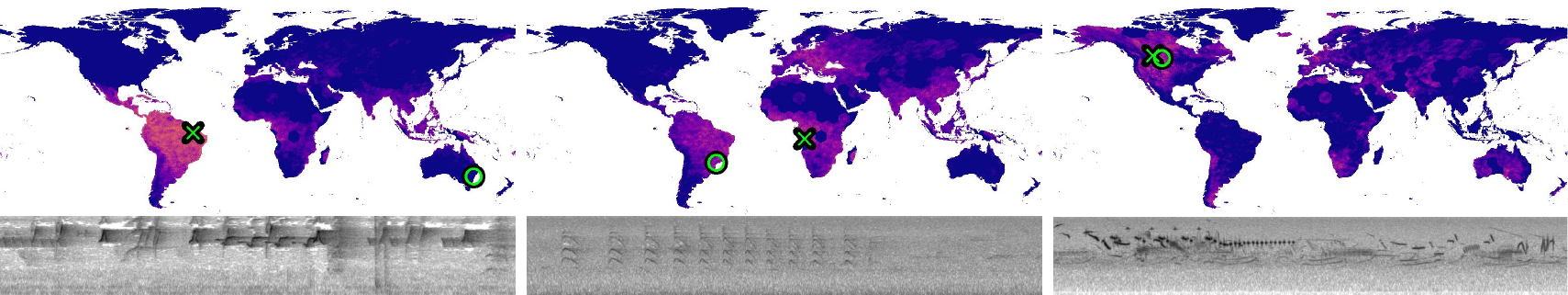}
    \begin{tabular}{ccc}
    \href{https://xeno-canto.org/442256}{Observation link} & \href{https://xeno-canto.org/442256}{Observation link} & \href{https://xeno-canto.org/277447}{Observation link}
    \end{tabular}
  \caption{
  \textbf{Model Predictions on XCDC.}  Heatmap shows the unnormalised likelihood of each location. Green crosses denote the final prediction (argmax) and green circles denote true location. Original observations linked below. 
  Best viewed zoomed in. 
  }
  \label{fig:predictions_supp_xcdc}
\end{figure}

See additional prediction visualizations of AG-CLIP on iNatSounds and XCDC in Fig~\ref{fig:predictions_supp_inat} and Fig~\ref{fig:predictions_supp_xcdc} respectively. These are a mix of success and failure cases of the model. 

\subsection{Classification Grids.}\label{sec:supp:classification}
\begin{figure}[ht]
  \centering
  \includegraphics[width=0.32\linewidth]{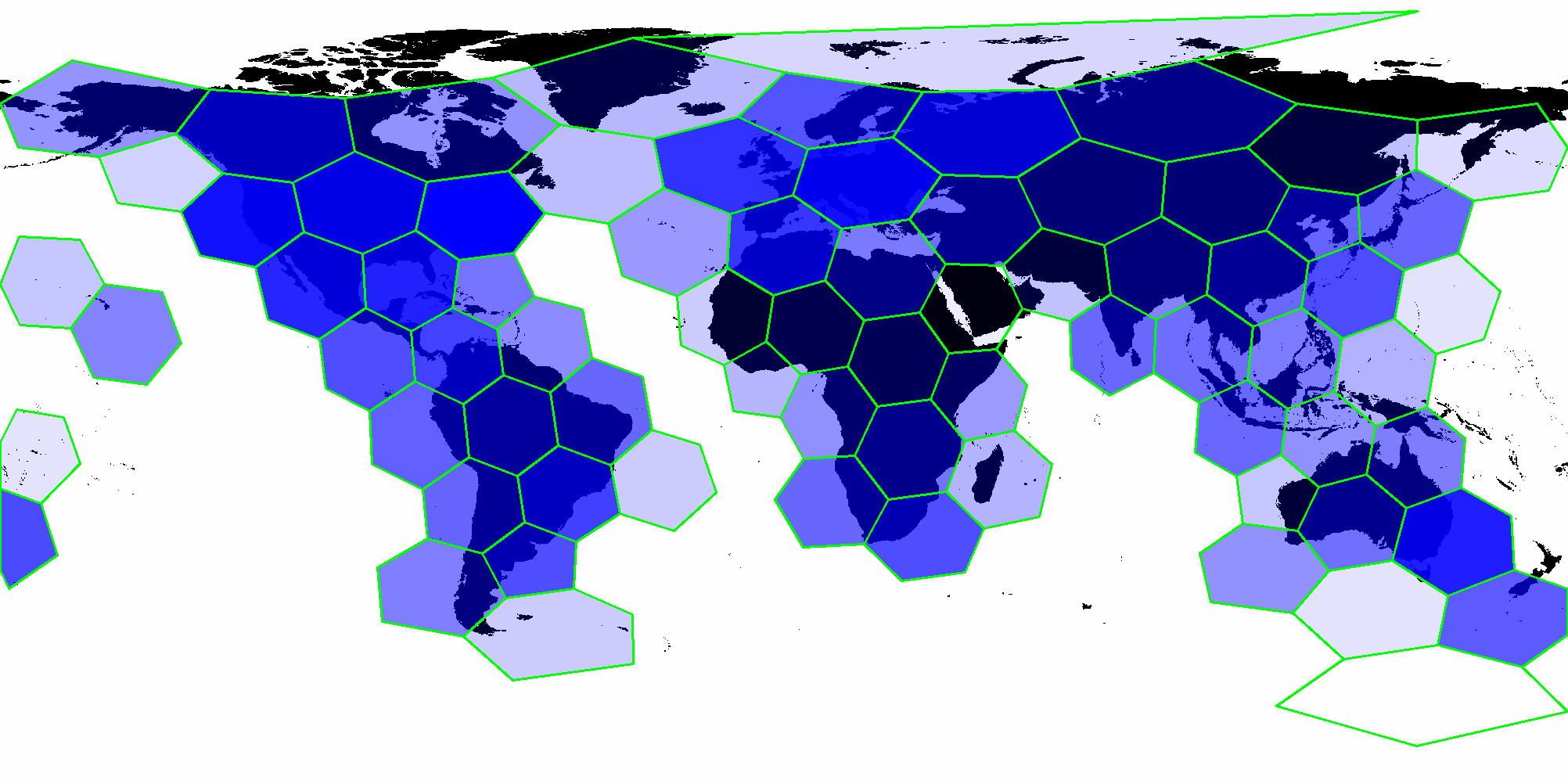}
  \includegraphics[width=0.32\linewidth]{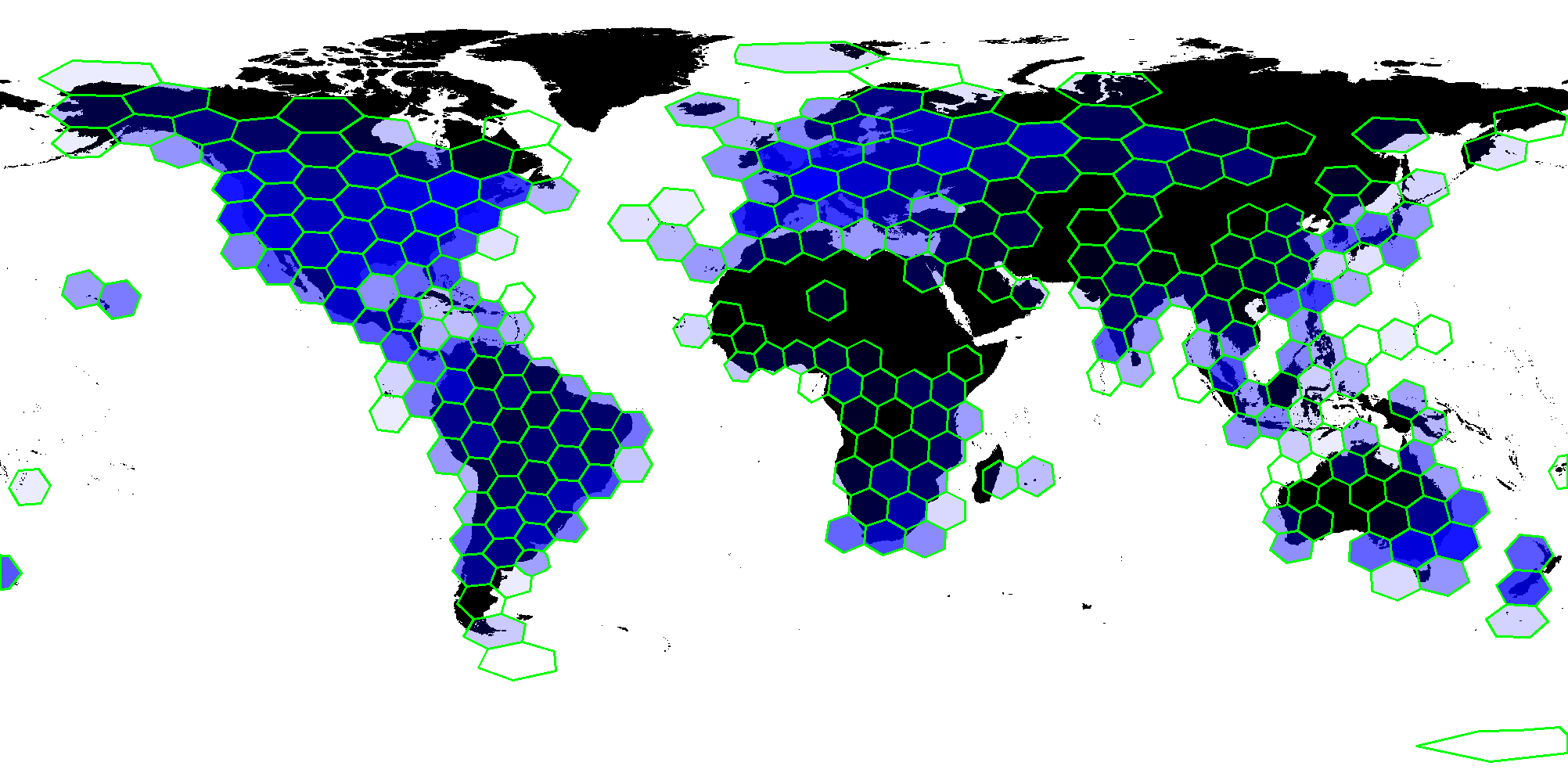}
  \includegraphics[width=0.32\linewidth]{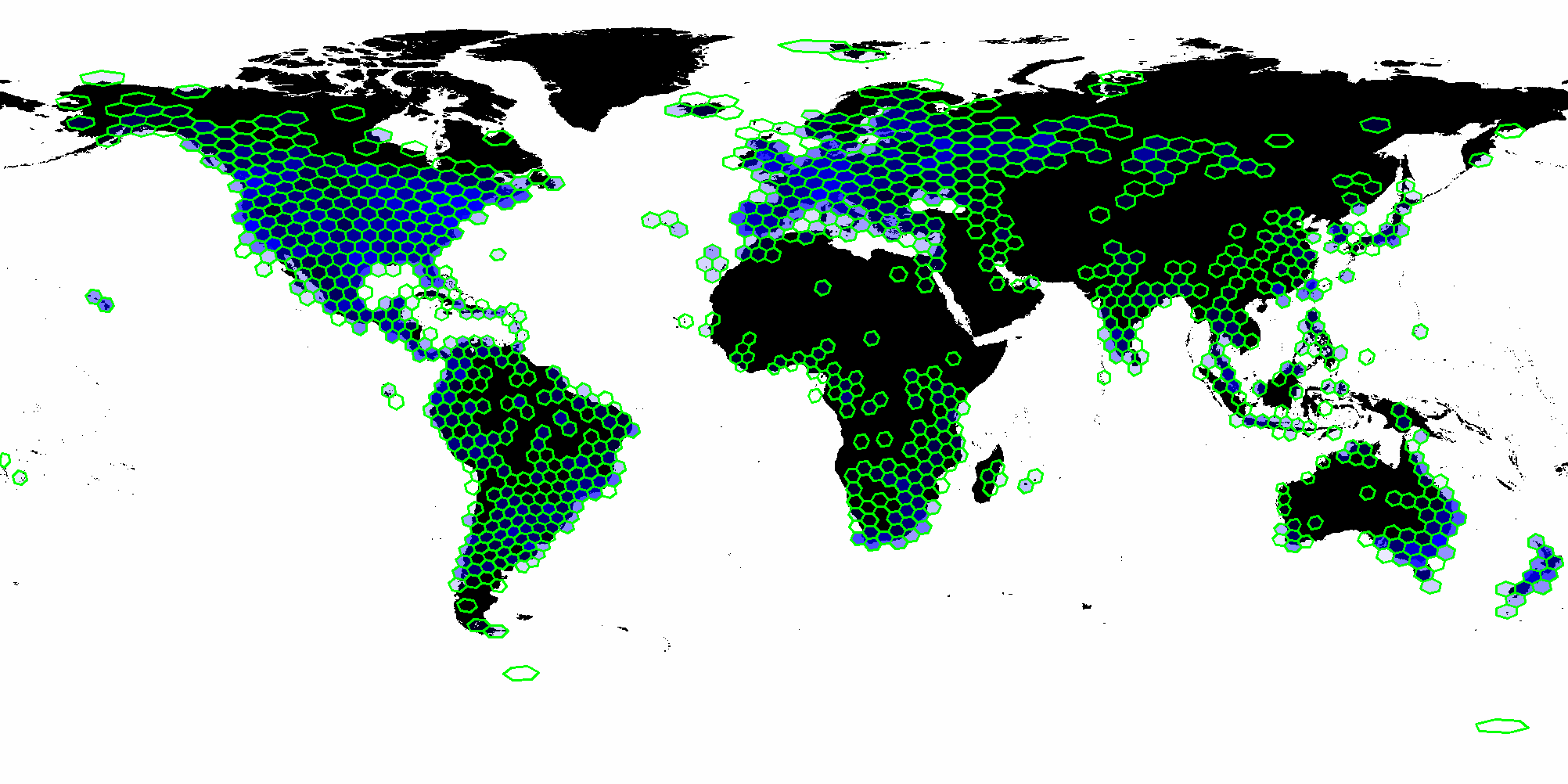}
  \caption{
  \textbf{Classification grids and iNatSounds Distribution.}
  Visualization of H3~\cite{brodsky2018h3} grid cells at different resolutions (0, 1 and 2 from left to right with average cell areas 436, 61 and 9 $\times 10^4 km^2$ respectively). In the hexagon itself, we show the actual distribution of the iNatSounds training set (higher opacity $\rightarrow$ more data). We use log scale for better visualization. 
  }
  \label{fig:class_grids}
\end{figure}

We visualize the different resolution hexagonal grids we use for classification in Fig~\ref{fig:class_grids}. We train the model to predict a hexagon from the list of all possible hexagons at a particular resolution. At test time, we use the center of the predicted hexagon as the geolocation prediction. As we go to finer resolutions, the ability of a classifier to pinpoint to finer scales increases. At the same time, the total number of hexagons also increases, which reduces classification performance.

We also show the distribution of iNatSounds training set on the same plots. The distribution is highly imbalanced, with North America and Europe having significantly higher number of recordings than other areas. 

\subsection{Retrieval Location Galleries.}\label{sec:supp:galleries}
\begin{figure}[t]
  \centering
  \begin{subfigure}[b]{0.32\textwidth}
      \includegraphics[width=\textwidth]{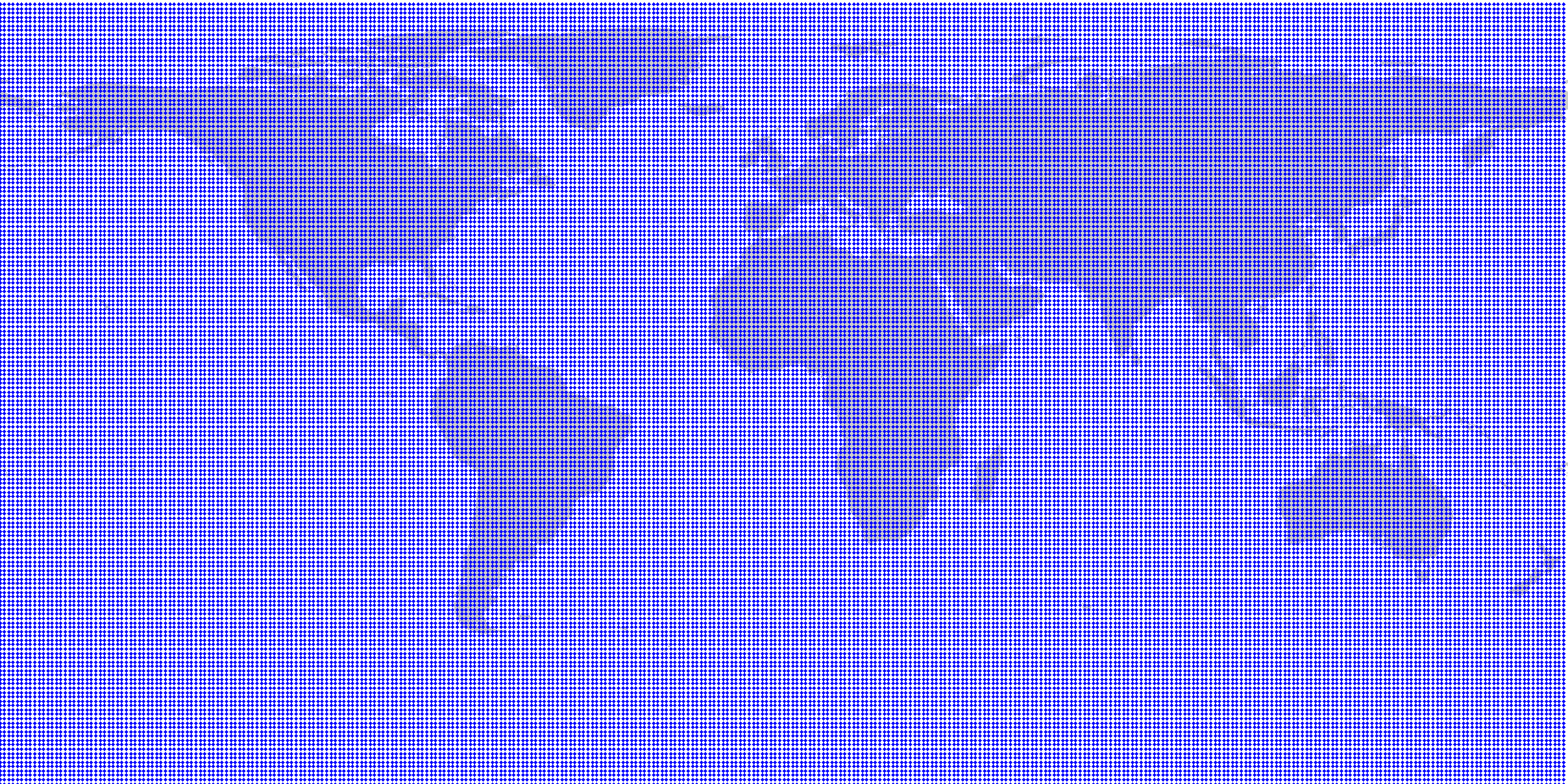}
      \caption{Uniform (65K points)}
  \end{subfigure}
  \begin{subfigure}[b]{0.32\textwidth}
      \includegraphics[width=\textwidth]{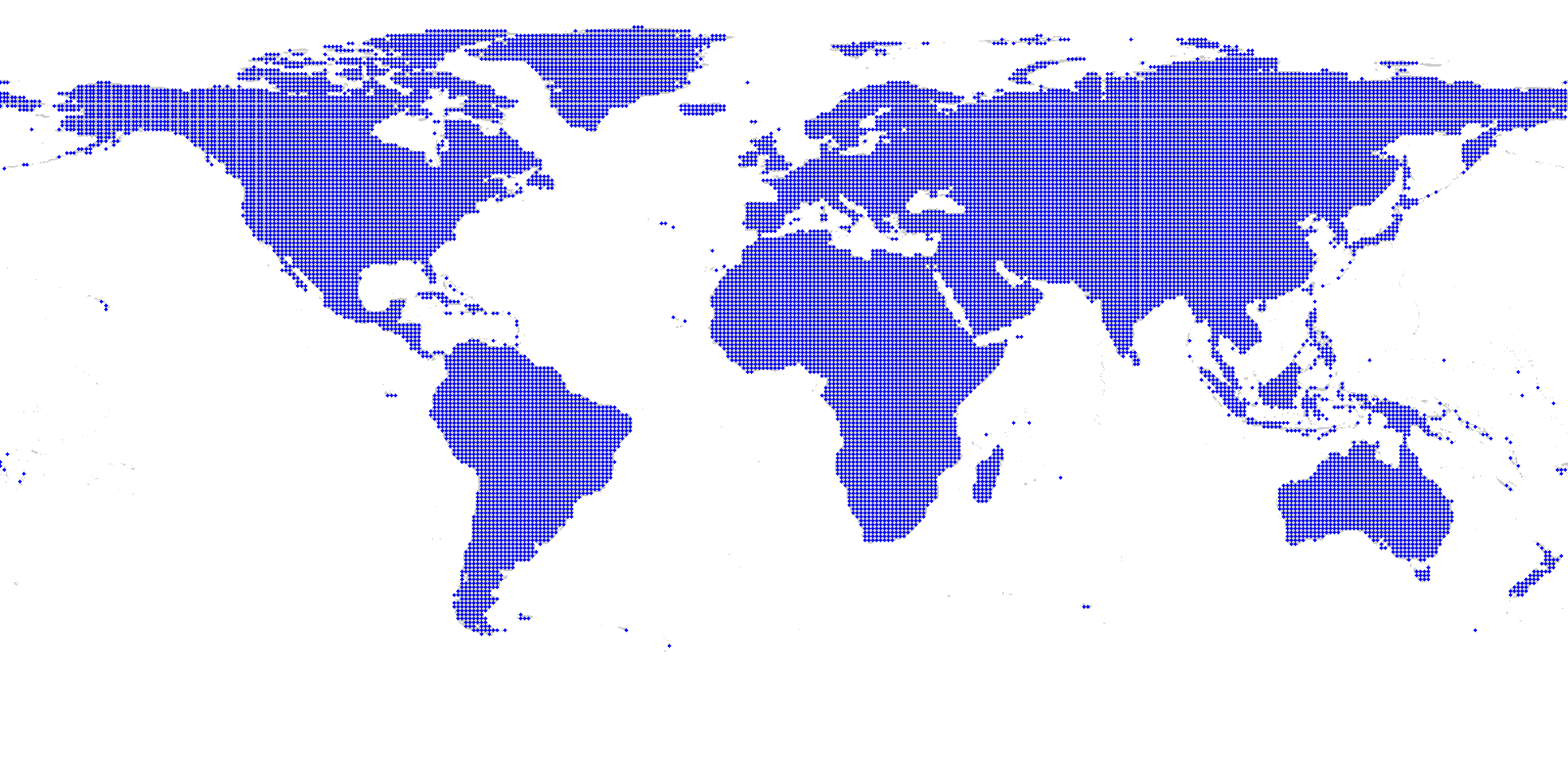}
      \caption{Land (20K points)}
  \end{subfigure}
  \begin{subfigure}[b]{0.32\textwidth}
      \includegraphics[width=\textwidth]{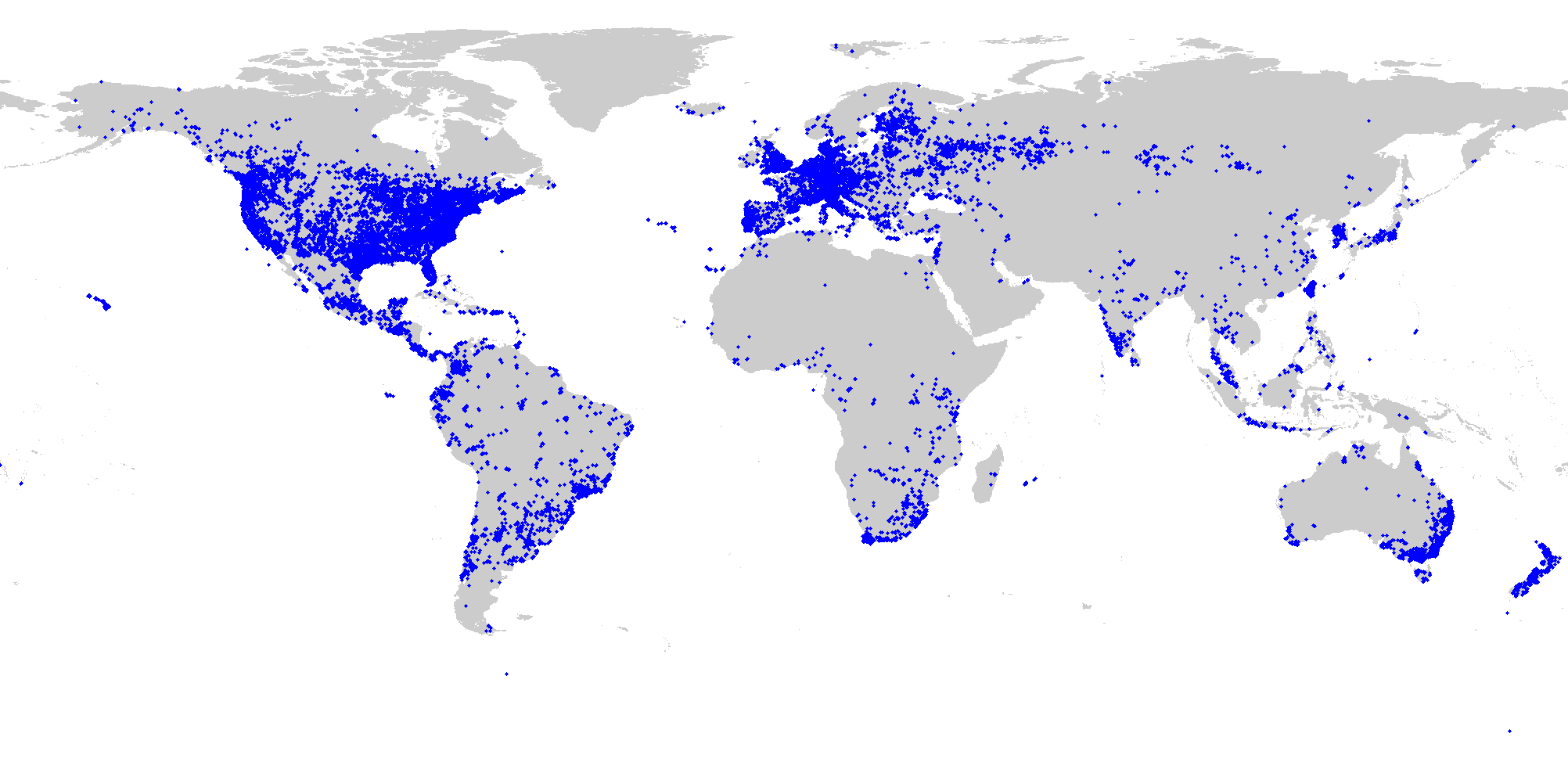}
      \caption{iNatSounds Train (120K points)}
  \end{subfigure}
  \begin{subfigure}[b]{0.32\textwidth}
      \includegraphics[width=\textwidth]{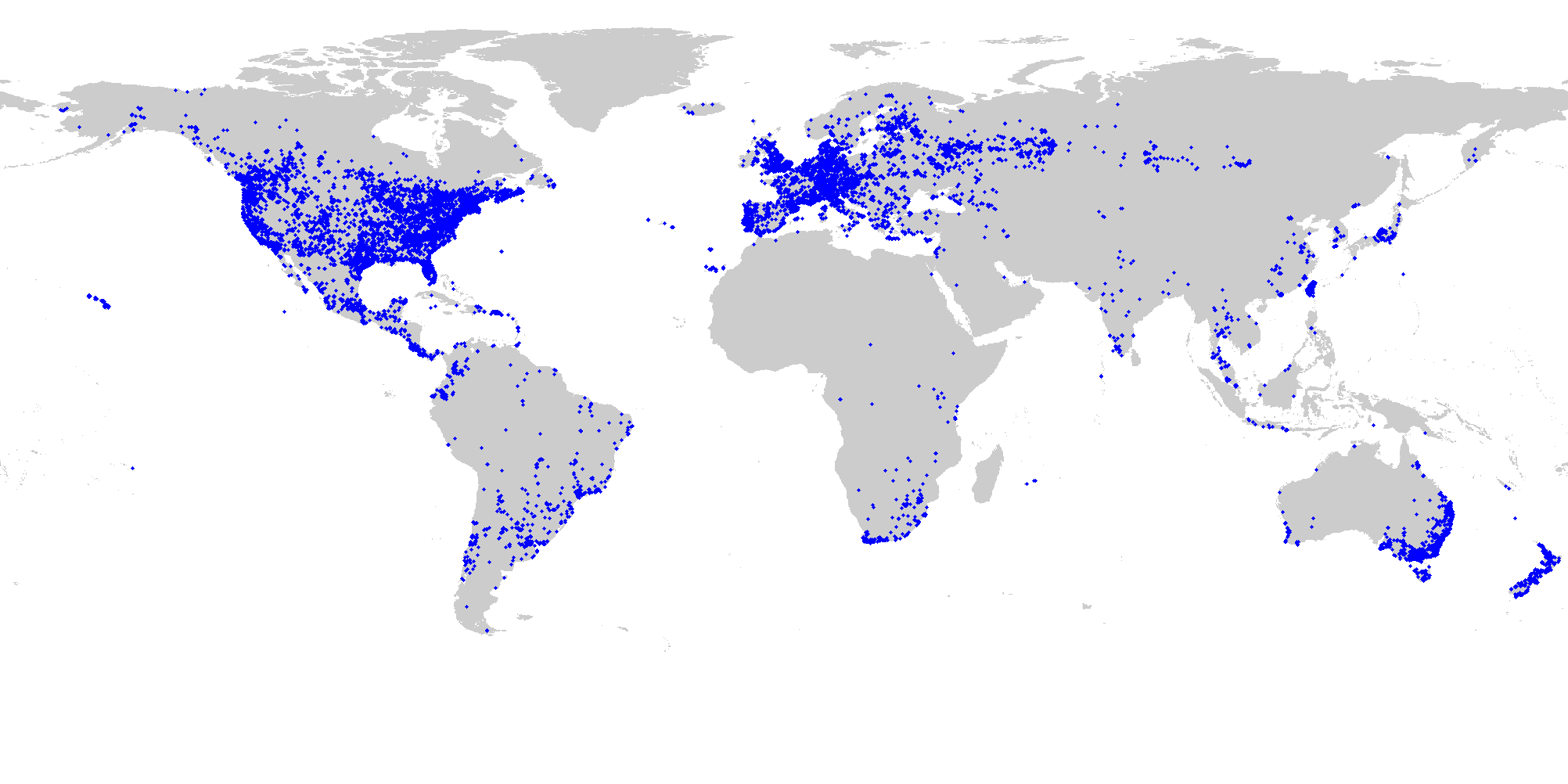}
      \caption{iNatSounds Val (45K points)}
  \end{subfigure}
  \begin{subfigure}[b]{0.32\textwidth}
      \includegraphics[width=\textwidth]{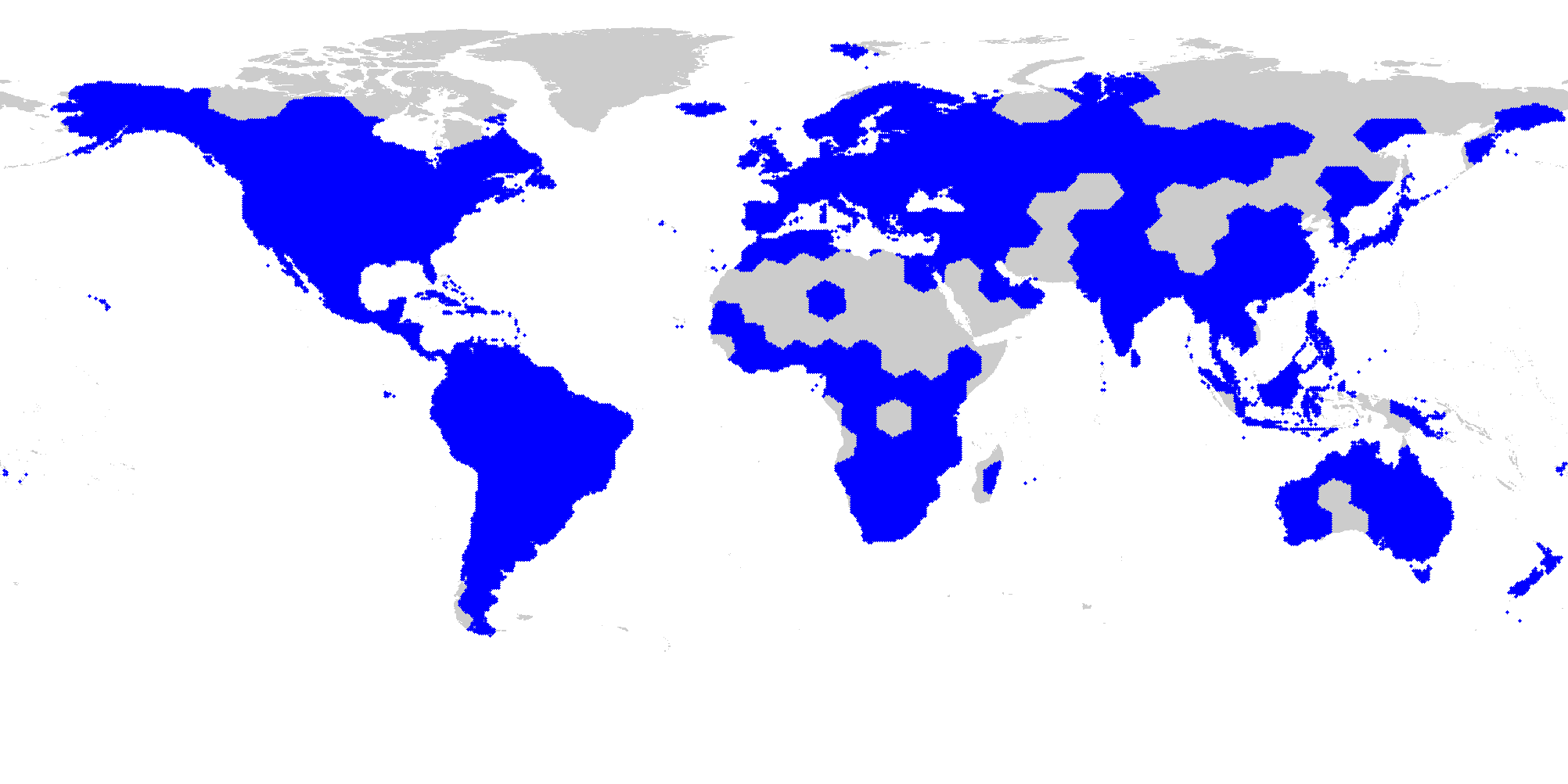}
      \caption{Train Neighbors (43K points)}
  \end{subfigure}
  \begin{subfigure}[b]{0.32\textwidth}
      \includegraphics[width=\textwidth]{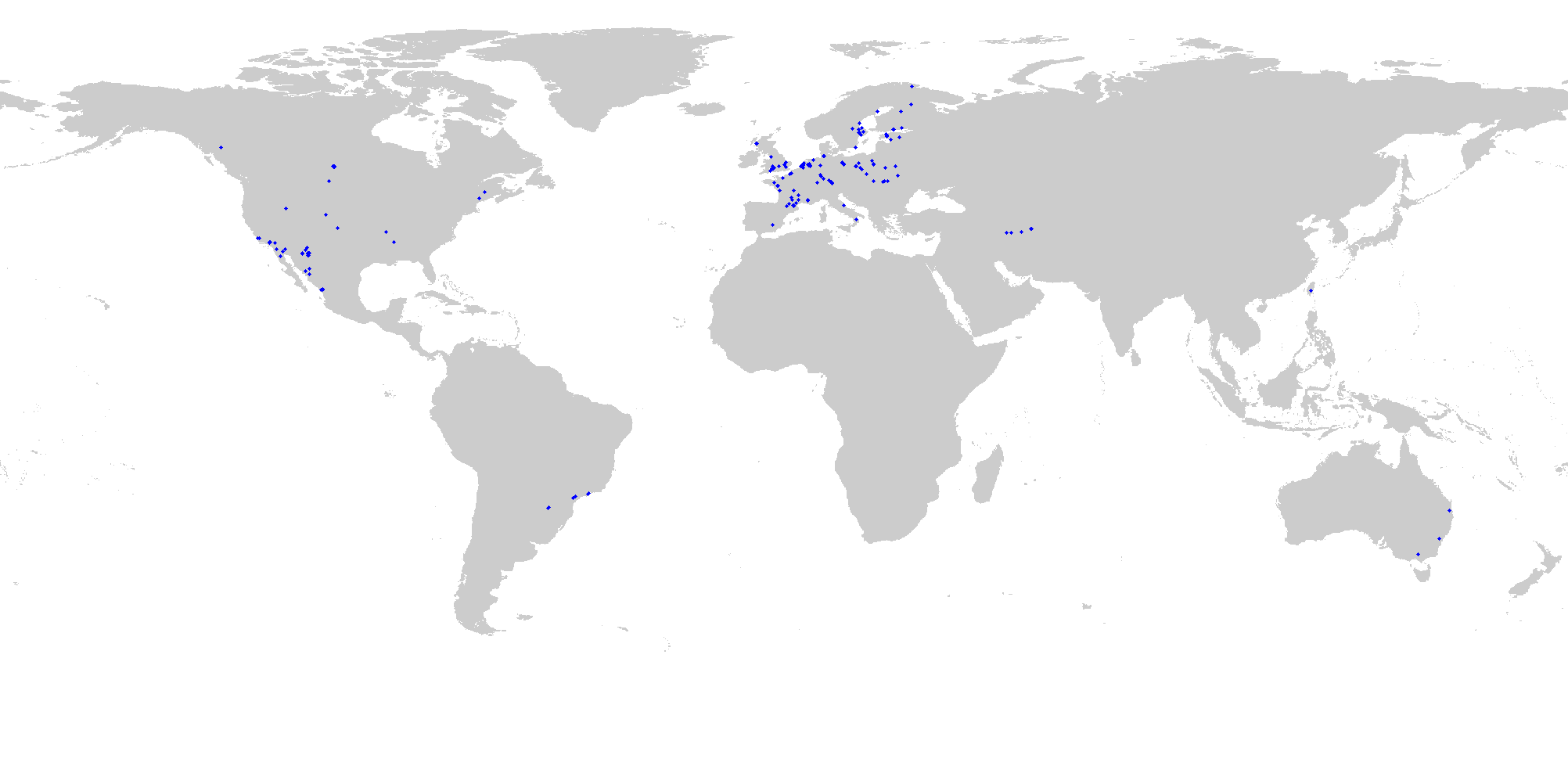}
      \caption{XCDC (579 points)}
  \end{subfigure}
  \caption{\textbf{Location Galleries.} Visualization of different galleries for evaluating geolocation. 
  }
  \label{fig:location_galleries}
\end{figure}

Location retrieval approaches like ours require a gallery of candidate locations to retrieve from at test time. An advantage of this approach over image retrieval approaches is that candidate locations can be sampled readily (just 2 coordinates) and do not require a database at test time. 
However, construction of these location galleries can affect final performance. A gallery that is too sparse can lead to higher geolocation errors and a gallery that is too dense can require too much compute. See Fig~\ref{fig:location_galleries} for visualization of  different galleries.

\begin{figure}[t]
  \centering
  \includegraphics[width=\textwidth]{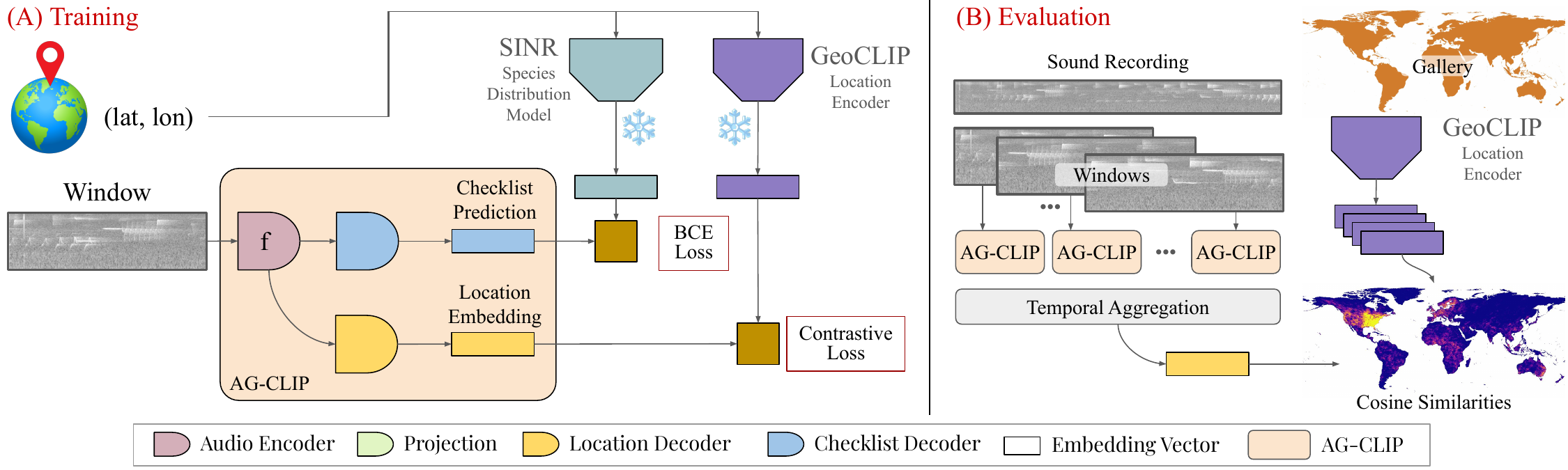}
  \caption{
  \textbf{Method Overview.}  
  At training time (left), we sample a window uniformly at random from each audio recording. Features extracted by the audio encoder $f_{\theta}$ are fed to a checklist decoder which is used in a BCE loss with checklists given by SINR~\cite{cole2023spatial} as labels.
  Audio features $f_{\theta}$ are also fed to a location decoder, whose outputs are matched with corresponding GeoCLIP~\cite{vivanco2024geoclip} location embeddings to get a constrastive loss. 
  The full model is trained with the BCE loss and Contrastive loss. At test time (right), we densely sample all windows from a recording, predict final location features via the full AG-CLIP model and then aggregate these features across windows to get a recording-level prediction. Finally, we compute the similarity of this feature with GeoCLIP embeddings of a gallery of candidate locations to pick the most similar, which is the geolocation prediction. 
  }
  \label{fig:supp_method}
\end{figure}

\subsection{Additional Implementation Details}
\label{sec:supp:implement}

\paragraph{Spectrogram Creation.}
We take a vision approach where 1D waveforms are converted to 2D spectrograms, which can be treated as images. 
Following previous work~\cite{chasmai2024inaturalist}, we generate spectrograms using the Short-Time Fourier transform (STFT),  with a window size of 512 and a stride length of 128. 
Linear spaced frequencies are converted to the mel-scale~\cite{pedersen1965mel}, mapping frequencies in the range [50Hz, 11.025kHz] to 128 logarithmically spaced mel bins, better aligning with human perception of pitch change. 
Each audio recording is split into a set of windows of 3 seconds each, strided by 1.5 seconds. Each window is treated as an independent image, repeated thrice and resized to get a $224 \times 224 \times 3$ RGB input.

\shortparagraph{Regression for Geolocation.}
An intuitive approach is to directly regress latitude and longitude from $f_{\mathbf{\theta}}(\mathbf{x})$. 
$h_{\mathbf{\phi}}$ can be a linear layer with 2 outputs and all weights can be trained using either Euclidean distance or the differentiable Haversine distance as the loss. 
While the former offers simplicity, the latter is more accurate as we are computing distances on a sphere instead of a plane. The (lat, lon) coordinates are normalised to  $[-1, 1]$ before being used as labels for regression. 
A third approach is to convert the spherical coordinates $(\text{lat}, \text{lon})$ to Cartesian coordinates $(x, y, z)$ and then use Euclidean distance~\cite{perotin2019regression}.

\shortparagraph{Classification for Geolocation.}
We can formulate geolocation as classification over a set of location bins. If the world is divided into bins, then $h_{\mathbf{\phi}}$ can be trained using cross-entropy to predict the bin that contains the recording location.

At test time, $h_{\mathbf{\phi}}$ predicts a probability distribution over the bins and we use the center of the most likely bin as the predicted location. 
The bin resolution (large bin area vs small bin area) imposes constraints on the maximum achievable performance since even correctly classified (\ie correctly binned) samples may be far from the bin center. This is likely the cause of the very poor regional performance with lowest resolution models in Table 1 (main paper). For higher resolution cells, the size of each cell is smaller, and thus, the error introduced by choosing the grid center is lower. However, higher resolution grids have more cells, or classes, which makes the classification itself somewhat harder.

We also experiment with a hierarchical approach to better handle this tradeoff. A low resolution model first predicts a large cell. Then a higher resolution model is used to predict a smaller cell within this large cell. This can be repeated for multiple levels of hierarchy. We start with resolution 0, followed by resolution 1 and finally 2. Note that this is done only at test time, so once the models at different resolution are trained, they can just be used directly for hierarchical classification.

\paragraph{Model Architecture and Hyperparameters.}
We show a block diagram of the AG-CLIP model architecture in Fig.~\ref{fig:supp_method}. The pipeline has some differences at training and test times. While training, we sample windows uniformly at random from audio recordings, while at test time, we densely sample strided windows from the entire audio recording and aggregate to get recording level predictions. 

In all experiments, we use the relatively light-weight MobileNet V3~\cite{howard2019searching} as our audio backbone to get a 1280 dimensional features from windows resized to 224 $\times$ 224. 
We choose  model dimension hyperparameters to get good performance while keeping model size comparable. 
Our checklist prediction head learns to predict the entire species checklist given by SINR. 
The projection layer takes the 1280 dimensional embeddings and projects it to the checklist dimension. 
The checklist and location decoders are both 2 layer MLPs with hidden dimensions of 128. 
After these modifications, the increase in model parameters is modest, from 4.4M for GeoCLIP to 4.7M for AG-CLIP.

We initialize the audio backbones with pretrained weights for iNatSounds~\cite{chasmai2024inaturalist} species identification. 
We take a weighted sum of the BCE and contrastive losses, with a weight of 0.01 on the BCE loss, which reflects its magnitude relative to the contrastive loss. 
We train for 50 epochs with early stopping using the Nesterov accelerated SGD optimizer with a batch size of 128, weight decay of 10$^{-5}$, and learning rate linearly ramped from 10$^{-3}$ to 10$^{-2}$ over the first 5 epochs and then cosine decayed back to 10$^{-3}$ over the remaining 45 epochs. These hyperparameters were chosen with the validation set released by iNatSounds. We report test set performance. 

\paragraph{Baselines: GeoCLAP and TaxaBind.}
For GeoCLAP, we get aerial imagery from the location with Google Maps API (similar to SoundingEarth~\cite{heidler2023self}), and use its image embedding as the location embedding in our retrieval setup. Instead of encoding the (lat, lon) coordinates directly with a model, we first get the corresponding aerial image, and embed this image with their image encoder. SoundingEarth, the dataset used by GeoCLAP also contains Google Maps images, and we match the zoom levels and image sizes with that dataset. We use their audio and image encoders off the shelf. 

TaxaBind includes encoders for 6 different modalities, all embedding in the same shared space. We use their audio and location encoders directly, in a setup very similar to ours. Note that TaxaBind does not actually use geotagged audio. It uses geotagged images and audio-image paired data, which are used to train location-image and audio-image encoders, respectively. Since all modalities share the same space, this training strategy allows us to do audio$\rightarrow$location retrieval, which we use for geolocation.

\subsection{Transformer to handle Variable Length}\label{sec:supp:transformer}

Models that can reason over time can potentially capture more interesting patterns. For longer recordings like those of XCDC, temporal analysis can allow a model to properly combine geographic cues from different species. Our default average pooling does this too, by way of a simple voting of different windows. We also explore if we can do better with a transformer.

We still remain in the late-fusion regime: geolocation features are predicted for each window independently and the transformer aggregates these predictions at the end. We do this with the training set so that the transformer takes in a sequence of predicted location features $N\times L$ and then predicts a single $L$-dimensional feature, where $L$ is the embedding dimension of the location encoder, which in our case is 512 for GeoCLIP. For easier batch construction, we keep the sequence length $N$ fixed at 32, padding or cropping appropriately. Note that this corresponds to around 50s of audio. We use 6 transformer layers, each with 8 heads and a hidden dimension of 128. Only the transformer layers are trained and the CNN backbone is kept frozen. We use Adam with learning rate of $10^{-3}$, weight decay of $10^{-3}$, and train for 50 epochs. 

\subsection{Ablation on Location Galleries}
\label{sec:supp:ablation}

\begin{table}[t]
    \centering
    \caption{\textbf{Location Galleries.} Effects of the location gallery used for AG-CLIP. The same AG-CLIP model (pre-trained on iNatSounds) is given different galleries at test time to get these results.}
    \label{tab:galleries}
    \setlength{\tabcolsep}{5pt}
    \begin{tabular}{lccccc}
        \toprule
        \multirow{2}{*}{\textbf{Gallery}} & \multirow{2}{*}{\textbf{\# Locs}} & \textbf{City} & \textbf{Region} & \textbf{Country} & \textbf{Continent}\\
        &  & 25km & 200km & 750km & 2500km \\
        \midrule
        Uniform & 65K & 01.3 & 15.0 & 38.6 & 69.2\\
        Uniform & 260K & 02.9 & 15.8 & 38.9 & 69.5\\
        Land & 20K & 01.6 & 15.4 & 38.6 & 69.6\\
        Train  & 137K  & 06.6 & 17.4 & 41.0 & 70.3\\
        Train Neighbors  & 43K  & 03.1 & 16.1 & 39.3 & 69.9\\
        Validation  & 45K  & 06.7 & 17.6 & 41.1 & 70.6\\
        XCDC  & 576  & 00.7 & 07.4 & 28.7 & 65.7\\
        \bottomrule
    \end{tabular}
\end{table}

We present ablations on the location gallery in Table~\ref{tab:galleries}. 
We first experiment with galleries constructed by uniformly sampling points on the 2D world map (latitude, longitude). We sample locations in intervals of either 1$^{\circ}$, leading to 65K locations, or 0.5$^{\circ}$, leading to 260K locations. We obtain similar performances for both, with region level at 13.4\% and 14.8\%, respectively. This indicates that adding additional locations on a uniform grid may not improve performance much, but will lead to much higher compute. Next, we again sample uniformly (1$^{\circ}$ interval), but restrict locations to only land mass. Since the earth is only about 29\% land, we are left with 20K locations, but performance does not change much, and actually improves slightly for some levels (city, region, and continent). This hints at the potential of smaller, but more targeted galleries.

Next, we use the train and validation sets to construct the gallery. Keeping the locations from each geo-tagged audio in the train set, we are essentially following the distribution of the dataset itself. Regions where recordings are dense in the training set have more locations in the gallery and conversely, sparse regions in the train set are allocated fewer gallery locations. If the test location distribution is similar to the train, this would be a good gallery, well balanced between granularity and compute cost. We could use the validation set instead to construct the gallery, although the train set is a more natural choice. Compared to the land gallery, we see small improvements at the continent and country level, but much bigger improvements at the region (16.4\% vs 13.8\%) and city (6.5\% vs 1.3\%) levels. The train neighbors gallery consists of uniformly sampled points like the uniform gallery, but only from H3~\cite{brodsky2018h3} hexagons that contain at least one training recording. This allows denser sampling with a better control over the total number of locations. Even with 0.5$^{\circ}$ intervals, the number of locations drops to 43K, at the cost of slightly worse performance.

\begin{table*}[t]
    \centering
    \caption{\small\textbf{Geolocation on XCDC}.  Models trained on iNatSounds training set and evaluated on XCDC.
    }
    \label{tab:supp:xcdc}
    \setlength{\tabcolsep}{4pt}
        \centering
        \small
        \begin{tabular}{clrcccc}
            \toprule
             \multicolumn{2}{c}{\multirow{2}{*}{\textbf{Experiment}}}  & \multicolumn{1}{c}{\textbf{$\downarrow$ Median}} & \textbf{$\uparrow$ City} & \textbf{Region} & \textbf{Country} & \textbf{Continent}\\
             &   & \multicolumn{1}{c}{\textbf{Error (km)}} & 25km & 200km & 750km & 2500km \\
            \midrule
            Naive & \multicolumn{1}{l}{Rnd Train Loc} & 9915 $\textcolor{stdgray}{\pm\ 99}$ & 00.0 $\textcolor{stdgray}{\pm\ 0.0}$ & 00.1 $\textcolor{stdgray}{\pm\ 0.2}$ & 00.5 $\textcolor{stdgray}{\pm\ 0.4}$ & 03.6 $\textcolor{stdgray}{\pm\ 0.5}$ \\
            \midrule
            \multirow{2}{*}{\shortstack[c]{Species\\Ranges}} & True Species & 449 $\textcolor{stdgray}{\pm\ 06}$ & 00.5 $\textcolor{stdgray}{\pm\ 0.2}$ & 25.8 $\textcolor{stdgray}{\pm\ 1.4}$ & 68.8 $\textcolor{stdgray}{\pm\ 1.4}$ & 99.0 $\textcolor{stdgray}{\pm\ 0.1}$ \\
             & Predicted Sp (All) &  1097 $\textcolor{stdgray}{\pm\ 18}$ & 00.1 $\textcolor{stdgray}{\pm\ 0.1}$ & 02.5 $\textcolor{stdgray}{\pm\ 0.3}$ & 27.6 $\textcolor{stdgray}{\pm\ 1.1}$ & 80.0 $\textcolor{stdgray}{\pm\ 0.3}$ \\
            \midrule
            \multirow{1}{*}{Classify}  & Hierarchical & 1116 $\textcolor{stdgray}{\pm\ 60}$ & 00.0 $\textcolor{stdgray}{\pm\ 0.0}$ & 05.3 $\textcolor{stdgray}{\pm\ 0.1}$ & 31.7 $\textcolor{stdgray}{\pm\ 1.9}$ & 69.7 $\textcolor{stdgray}{\pm\ 0.6}$ \\
              \midrule
             \multirow{2}{*}{Retrieve} & AG-CLIP (\textbf{ours}) & 1112 $\textcolor{stdgray}{\pm\ 93}$ & 00.2 $\textcolor{stdgray}{\pm\ 0.2}$ & 04.3 $\textcolor{stdgray}{\pm\ 0.7}$ & 26.3 $\textcolor{stdgray}{\pm\ 3.5}$ & 71.9 $\textcolor{stdgray}{\pm\ 0.2}$\\
             & AG-CLIP (XCDC gallery) & 912 $\textcolor{stdgray}{\pm\ 72}$ & 05.9 $\textcolor{stdgray}{\pm\ 0.2}$ & 11.6 $\textcolor{stdgray}{\pm\ 0.7}$ & 36.6 $\textcolor{stdgray}{\pm\ 3.1}$ & 73.8  $\textcolor{stdgray}{\pm\ 0.3}$\\
            \bottomrule
        \end{tabular}%
\end{table*}

\subsection{Audio Geolocation on XCDC}
\label{sec:supp:xcdc}

We report repeated runs with mean and standard deviations for XCDC experiments in Table~\ref{tab:supp:xcdc}. Standard deviations are relatively low. We also include an experiment with AG-CLIP where we use the XCDC-gallery instead of iNatSounds. We see a good boost in performance, particularly at finer scales. However, these improved results are still worse that iNatSounds, supporting our hypothesis about the difficulty of our models to identify species rich data in XCDC.

\begin{table*}[t]
    \centering
    
    \caption{
     \small\textbf{Geolocation on WABAD}. Models trained on iNatSounds training set and evaluated on WABAD~\cite{perez2025wabad}. 
    }
    \label{tab:wabad}
    \setlength{\tabcolsep}{7pt}
    \centering
    \small
    \begin{tabular}{clrcccc}
    \toprule
    \multicolumn{2}{c}{\multirow{2}{*}{\textbf{Experiment}}}  & \multicolumn{1}{c}{\textbf{$\downarrow$ Median}} & \textbf{$\uparrow$ City} & \textbf{Region} & \textbf{Country} & \textbf{Continent}\\
    &   & \multicolumn{1}{c}{\textbf{Error (km)}} & 25km & 200km & 750km & 2500km \\
    \midrule
    \multirow{2}{*}{Regress} & Euclidean & 3540 & 0.00  & 0.80  & 07.51  & 37.77 \\
    & Haversine &  3853 & 0.00  & 0.86  & 08.53  & 37.66 \\
    \midrule
    \multirow{3}{*}{Classify} &  Res-0 (430 $\times 10^4$km$^2$) & 4126 & 0.00 & 0.00 & 09.48 & 40.38 \\
    & Res-2 (8.6 $\times 10^4$km$^2$) & 5404 & 0.00 & 1.44 & 08.93 & 34.50 \\
    & Hierarchical (0$\rightarrow$1$\rightarrow$2) & 4746 & 0.00 & 1.30 & 12.39 & 36.42 \\
    \midrule
    \multirow{3}{*}{\shortstack[c]{Species \\Ranges}} & Annotated Species & 736 & 0.60 & 17.32 & 50.33 & 79.59\\
    & Predicted Species (Top 1) & 9678 & 0.00 & 0.37 & 01.63 & 07.23 \\
    & Predicted Species (All) & 3317 & 0.35 & 2.53 & 15.69 & 44.00\\
    \midrule
    \multirow{3}{*}{Retrieve} & GeoCLAP & 7139 & 0.10 & 0.58 & 02.91 & 15.20 \\
    & Taxabind & 7429 & 0.10 & 0.44 & 04.46 & 15.81 \\
    & AG-CLIP (\textbf{ours}) & 3439 & 0.20 & 3.23 & 14.90 & 45.03 \\
    \bottomrule
    \end{tabular}

\end{table*}

\subsection{In-the-wild Audio Geolocation with WABAD}
\label{sec:supp:wabad}

Our models are trained with iNatSounds, characterized by short, focal recordings. In contrast, passive acoustic monitoring~\cite{sugai2019terrestrial} (PAM) projects often collect longer recordings that capture not only target species vocalizations but also background ambient sounds. This domain shift from the training distribution presents an opportunity to test our models on challenging and realistic in-the-wild settings.

WABAD~\cite{perez2025wabad} is one such PAM dataset with dense annotations (timestamps and frequency ranges) covering \textbf{91K} animal vocalization from more than \textbf{1.1K} species around the world. 
We present the performance of our models on this dataset in Table~\ref{tab:wabad}. Similar to XCDC, we see significant drops compared to the performance on iNatSounds, further highlighting the challenges of this domain shift. However, the relative performance of different methods is generally consistent (lower performance of classification being an exception). \method demonstrates better robustness than the soundscape mapping baselines Taxabind and GeoCLAP, particularly at the country and continental levels.

\subsection{Additional Details for Multimodal Geoforensics}
\label{sec:supp:geo_forensics}

We use Youtube clips for each movie shot we investigate. Searching for the moment where we suspect a discrepancy (based on viewer comments), we extract 10s audio clips and take a screenshot. The audio is geolocated by AG-CLIP. For the image, we use CLIP image encoder, followed by a (frozen) projection, as done by GeoCLIP. We visualize geolocation predictions of each modality of each example by arrows. These arrows show only the approximate location, but we compute the discrepancy errors exactly. 

GeoCLIP keeps their image encoder frozen and trains a location encoder. This allows them to swap in CLIP's text encoder instead of image, facilitating geolocation of text. Text descriptions of visual or acoustic scenes can serve as another modality for geoforensics.

\begin{figure}
    \centering
    \begin{subfigure}[b]{0.41\textwidth}
        \centering
        \includegraphics[width=\linewidth]{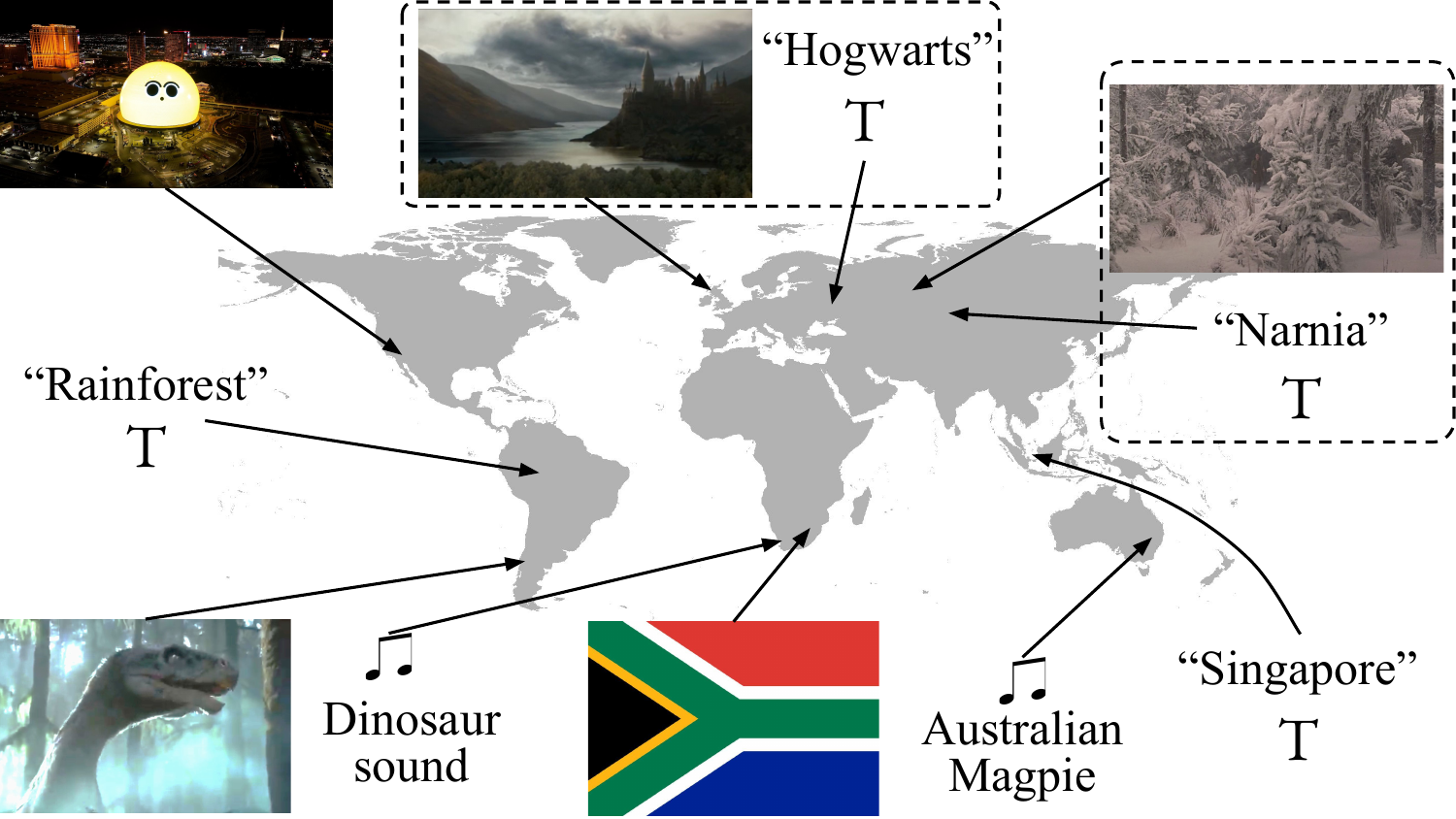}
        \caption{
          \textbf{Multimodal Geolocation.}
          Here, we explore geolocation using audio, images, and text. 
          We try out some fun experiments and see what the geolocation models predict for these inputs. For some of these prompts like ``Narnia'' or Dinosaur sounds, we do not know what the correct location should be, but it is interesting to see what the model predicts.
          }
          \label{fig:multimodal_geo}
    \end{subfigure}%
    \hfill
    \begin{subfigure}[b]{0.57\textwidth}
        \centering
        \includegraphics[width=\linewidth]{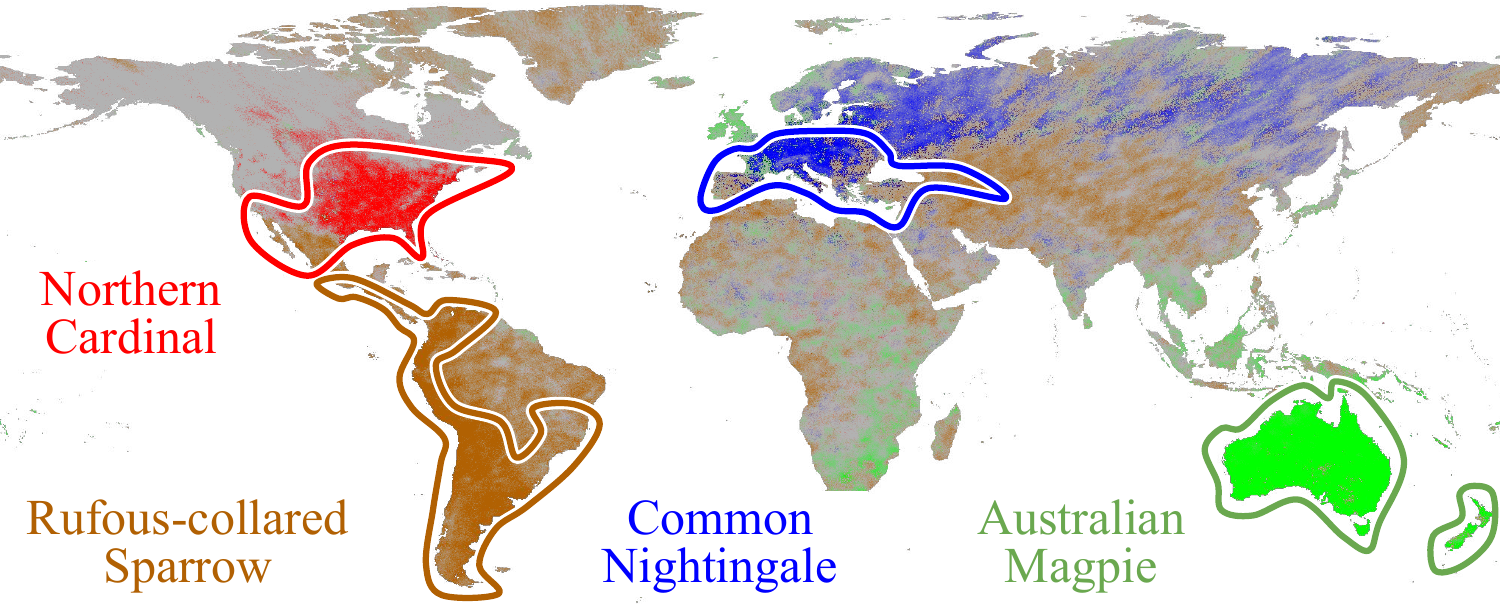}
        \caption{
          \textbf{Soundscape Affinities of Species.} 
          Heatmaps of cosine similarity between the average soundscape embedding of selected species and a gallery of location embeddings sampled globally. High similarity regions (darker areas) indicate locations with soundscapes that match the species’ vocal features, showing alignment with known ranges and highlighting unexpected affinities with acoustically similar environments beyond  typical habitats.
          }
          \label{fig:avg_predictions}
    \end{subfigure}%
    \caption{Additional experiments and visualizations}
\end{figure}

\subsection{Multimodal Geolocation}
\label{sec:supp:multimodal}

In \method, we keep the location encoder frozen from GeoCLIP, who in turn had kept their image encoder frozen from CLIP. 
Thus, we have a set of encoders that embed audio, location, images and text to the same space, allowing us to do multimodal geolocation (see Fig~\ref{fig:multimodal_geo}). Interestingly, CLIP can recognize flag images as well, and can geolocate them to the correct countries. 
The Las Vegas sphere (top left) finished construction after these models were trained, but they are interestingly still able to geolocate it accurately, perhaps because of some background details. 

Text helps expand the scope of queries in this problem. 
We can ask the model to geolocate the text ``rainforest'', and it predicts a location close to the Amazons, the biggest rainforest in the world. 
Giving country names as text queries also seems to work well. As fun experiments, we also ask the model to geolocate fictional places like ``Hogwarts'' and ``Narnia.'' For Narnia, the model predicts a location near Russia, which may be due to the snowy and mountainous landscapes seen in the first part of the movie. 
Image geolocation on a video frame from the movie also points to a similar location. 
Similarly for Hogwarts, the predicted locations of text and image geolocation models are interestingly close. 
While we cannot say whether the predicted geolocations for these are correct or not, it can be a fun way to probe what these models are learning.

\subsection{Soundscape Affinities of Species.}\label{sec:supp:affinity}

How does the soundscape associated with a species relate to its geographic range? We explore this question using AG-CLIP and visualize the results in Fig.~\ref{fig:avg_predictions}. For each species, we compute the average audio embedding using test recordings from iNatSounds associated with that species. We then calculate the cosine similarity between this average embedding and a gallery of location embeddings uniformly sampled across the globe (restricted to land locations). The similarities are clamped to the range [0, 1] and used as the alpha channel in the heatmap visualizations.

Fig.~\ref{fig:avg_predictions} shows results for four species: Northern Cardinal, Rufous-collared Sparrow, Common Nightingale, and Australian Magpie, alongside coarse approximations of their range maps. The soundscape embedding of the Northern Cardinal aligns well with its range, but interestingly, it does not capture the western portion of its range as strongly, possibly reflecting the distinct soundscapes of eastern forests versus western deserts. For the Rufous-collared Sparrow, the soundscape feature extends far beyond its range in South America, likely due to its occupancy of semiopen habitats (\ie villages, towns, and farmland), making it acoustically similar to many locations worldwide. The Common Nightingale’s soundscape affinity stretches into Russia, suggesting a similarity in soundscapes between northern Europe and parts of Russia. Finally, the Australian Magpie’s embedding aligns with its range in Australia and New Zealand but also shows unexpected affinities with deserts in Africa and rainforests in Southeast Asia, two habitats present in Australia.

These results highlight that the soundscape features learned by AG-CLIP capture both expected and surprising affinities of species soundscapes, providing insights into the relationship between species vocalizations and the broader acoustic environment.

\begin{figure}[t]
  \centering
  \includegraphics[width=\textwidth]{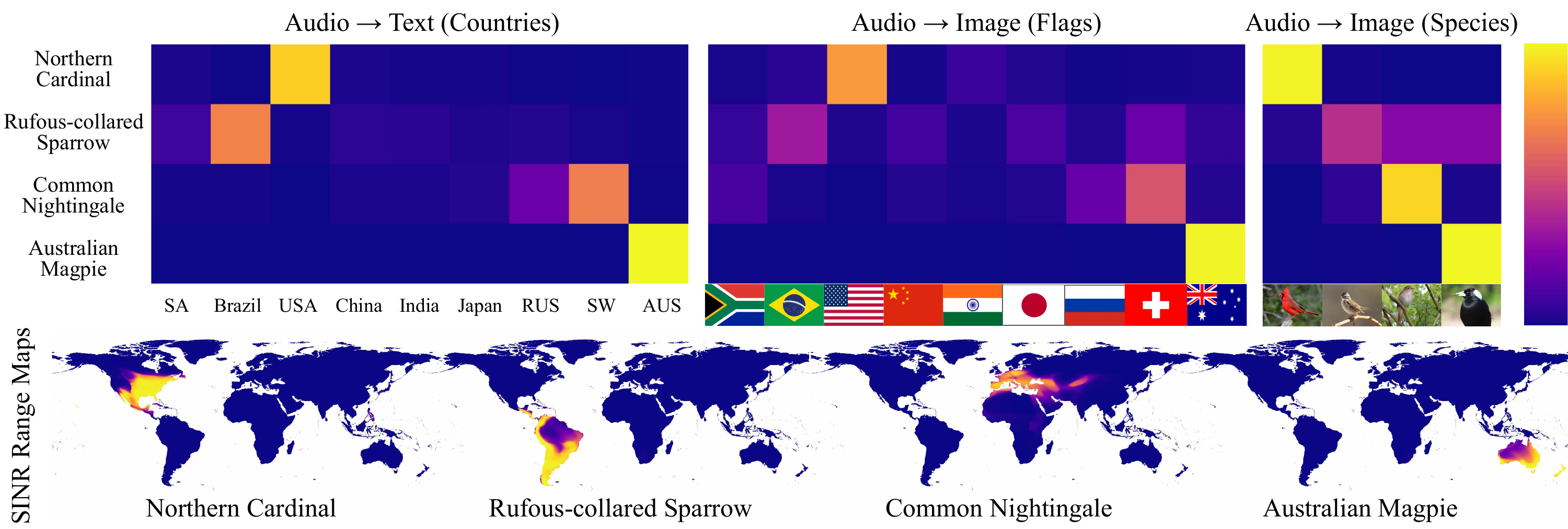}
  \caption{
  \textbf{Multimodal Retrieval.} 
  We visualize some examples for Audio $\rightarrow$ Text and Audio $\rightarrow$ Image retrieval. 
  We use the same average soundscape embeddings of species shown in Fig.~\ref{sec:supp:affinity} and retrieve some text or image queries. We show similarity matrices for 1) Country names as text, 2) Country flags as images and 3) Same species as images.
  Retrieval for country texts and flag images are relatively consistent. For species image retrieval, we see some confusion for the Rufous-collared sparrow, but all species are correctly matched. 
  At the bottom, we show the SINR range maps of these species for context of the countries these species are typically found in. 
  }
  \label{fig:visualizations}
\end{figure}

\subsection{Retrieving Images and Text from Audio}
\label{sec:supp:audio2image}

Since we embed audio in a shared text-image-location feature space, we can use our geolocation models to perform Audio $\rightarrow$ Text and Audio $\rightarrow$ Image retrieval by simply replacing location embeddings with the corresponding modality. We present a few such experiments in Fig~\ref{fig:visualizations}.

We use the same average soundscape embeddings of species shown in Fig.~\ref{sec:supp:affinity}. We compute similarities of these embeddings with certain text or image embeddings for retrieval. 
First we attempt Audio $\rightarrow$ Text by comparing each species feature with text features of country names. 
Notably, we observe some confusion between Switzerland and Russia for the Common Nightingale. These similarities closely reflect the SINR~\cite{cole2023spatial} species range maps for corresponding species seen at the bottom of the figure. 
Next, we attempt Audio $\rightarrow$ Image by using images of flags of the same set of countries. We observe predictions very similar to the Text retrieval experiment, which underscores how well these modalities are paired. 
Finally, we repeat the Audio $\rightarrow$ Image experiment by using species images instead of flags. While there is some confusion for the Rufous-collared Sparrow, the other three species are surprisingly confident and the model retrieves all species correctly. 
Some multimodal retrieval tasks are often constrained by the availability of paired data from each modality. These experiments show a promising and flexible alternative of using ``weakly'' paired data where some but not all modalities are paired with each other.

\subsection{Taxonomic Breakdown.}
\label{sec:supp:taxon_split}
Please see Table~\ref{tab:taxon_split} for the amount of data and geolocation results per taxonomic class.
It is dominated by birds in both the number of species and recordings. Amphibians and Mammals make up only 11.7\% and 5.3\% of the species, respectively.

\begin{table}[!h]
    \centering
    \small
    \caption{
    \textbf{Taxonomic breakdown of iNatSounds}.  Left: Number of species and recordings for  different taxonomic classes. Right: Geolocation performance for each taxonomic class.
    }
    \label{tab:taxon_split}
    \setlength{\tabcolsep}{4pt}
    \begin{minipage}{0.37\textwidth}
    \begin{tabular}{lrr}
        \toprule
        Class & \multicolumn{2}{c}{Train} \\
        & Species & Recordings\\
        \midrule
        Aves& 3,846 & 111,029 \\
        Insecta& 745 & 10,080  \\
        Amphibia& 650& 13,183  \\
        Mammalia&296& 2,566  \\
        Reptilia& 32 & 154 \\
        \bottomrule
            \end{tabular}
    \end{minipage}
    \begin{minipage}{0.61\textwidth}
        \begin{tabular}{lccccc}
        \toprule
        Class & Median & City & Region & Country & Continent\\ 
        &  Error (km) & 25km & 200km & 750km & 2500km \\
        \midrule
        Aves & 1165 & 5.9 & 15.68 & 38.79 & 69.81\\ 
        Insecta & 679 & 12.9 & 26.99 & 54.58 & 79.55\\ 
        Amphibia & 615 & 7.7 & 26.61 & 57.56 & 83.75\\ 
        Mammalia & 1442 & 3.6 & 11.84 & 33.27 & 61.32\\ 
        Reptilia & 2261 & 0.0 & 0.00 & 12.50 & 50.00\\ 
        \bottomrule
        \end{tabular}
    \end{minipage}
\end{table}

\subsection{Other Bioacoustic Audio Encoders.}
\label{sec:supp:bio_encoders}
The first row shows the full AG-CLIP results from Table~\ref{tab:main_neurips}. For a direct comparison, we also report linear-probe results using the same MobileNetV3 backbone pretrained on iNatSounds species identification. MobileNet features generally outperform TaxaBind~\cite{sastry2025taxabind}, BirdMAE~\cite{rauch2025can}, and ProtoCLR~\cite{moummad2026domain} across distance thresholds, likely because of the benefit of in-domain training, as those models are trained on Xeno-Canto or smaller iNaturalist datasets. Their bird-only training may also be a disadvantage, since iNatSounds includes amphibians, reptiles, insects, and mammals in addition to birds. Perch 2.0~\cite{van2025perch, chasmai2026metaperch} performs much better than the vanilla MobileNet linear probe, with a 16.5\% gain at the continent level. However, the training set of Perch might have some overlap with our evaluation set. Despite this, the fine-tuned MobileNet used in AG-CLIP remains the best overall, with consistent improvements over Perch 2.0 across spatial scales.

These results complement the ablations in Table 2a, which were intended to highlight the advantage of bioacoustic training over general audio backbones.

\begin{table}[ht]
    \centering
    \caption{\textbf{Other bioacoustic audio encoders.} Off-the-shelf audio encoder performance with retrieval-based linear probes on iNatSounds.}
    \label{tab:bio_encoders}
    \begin{tabular}{cccccc}
    \toprule
         Audio Encoder & Fine-tuned? & City (\%) & Reg. (\%) & Cou. (\%) & Con. (\%)\\
    \midrule
        MobileNetV3 (AG-CLIP) &  \checkmark & 6.4 & 17.2 & 41.0 & 71.2\\
        MobileNetV3 & \xmark & 3.3 & 9.6 & 26.7 & 54.4\\
        Taxabind & \xmark & 1.7 & 5.6 & 17.8 & 42.1\\
        BirdMAE & \xmark & 2.4 & 6.4 & 18.8 & 43.0\\
        ProtoCLR & \xmark & 2.5 & 7.5 & 23.0 & 50.6\\
        Perch 2.0 & \xmark & 5.1 & 15.0 & 38.9 & 70.9\\
    \bottomrule
    \end{tabular}
\end{table}

\subsection{Audio Encoder Architectures.}
\label{sec:supp:architectures}
Please see Table~\ref{tab:backbones} and Table~\ref{tab:resnet} for additional experiments with other audio encoder architectures.

\begin{table}[!h]
    \centering
    \caption{\textbf{Audio Backbones.} Experiments with other backbone architectures for the audio encoder. }
    \label{tab:backbones}
    \begin{tabular}{lccccc}
         \toprule
Audio Backbone & Median Geolocation & City & Region & Country & Continent \\
\midrule
MobileNet-V3 & 1082 & 6.40 & 17.2 & 41.0 & 71.2 \\
ResNet50 & 935 & 6.97 & 18.86 & 44.27 & 74.37 \\
ViT-B16 & 1069 & 7.10 & 17.79 & 41.19 & 71.24 \\
\bottomrule
    \end{tabular}
\end{table}

\begin{table}[h]
    \centering
    \caption{\textbf{Baselines with ResNet50.} Results for AG-CLIP and other methods for ResNet50. }
    \label{tab:resnet}
    \begin{tabular}{lccccc}
    \toprule
    ResNet50 experiment & Median Geolocation & City & Region & Country & Continent \\
    \midrule
    mse & 1650 & 0.05 & 2.67 & 23.75 & 63.05 \\
    classification (res=2) & 1114 & 0.36 & 18.10 & 40.22 & 69.22 \\
    Retrieval (SatCLIP) & 982 & 1.83 & 16.28 & 43.30 & 70.99 \\
    Retrieval (AG-CLIP) & 935 & 6.97 & 18.86 & 44.27 & 74.37 \\
    \bottomrule
    \end{tabular}
\end{table}

\subsection{Data Resampling.}
\label{sec:supp:data_resampling}
\begin{figure}[!h]
    \centering
    \includegraphics[width=\linewidth]{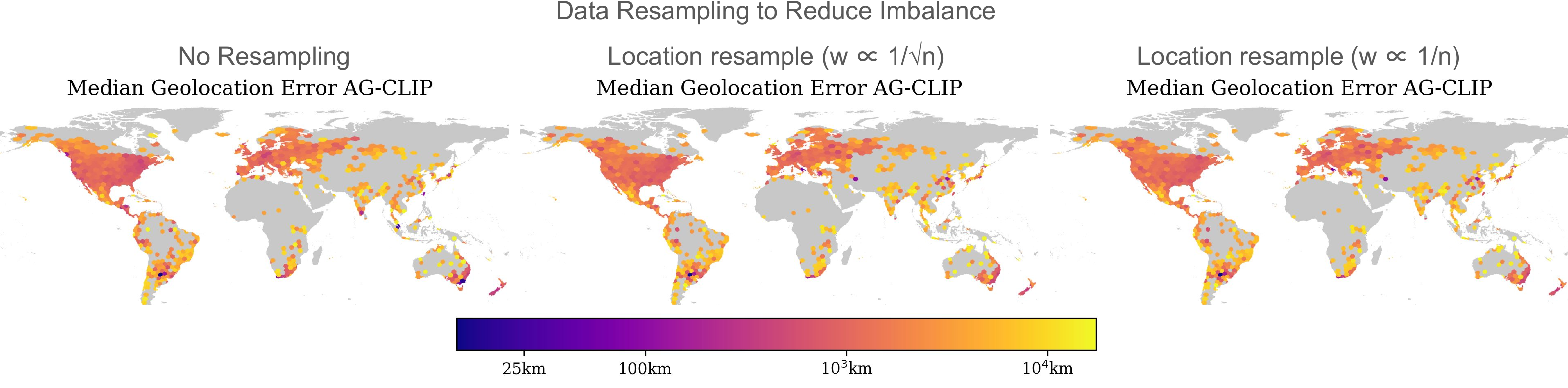}
    \caption{\textbf{Location weighted sampling.} Geographic distribution of performance for different data resampling strategies.}
    \label{fig:class_balancing}
\end{figure}

Please see Table~\ref{tab:resampling} and Fig~\ref{fig:class_balancing} for the experiments with data resampling to mitigate the effects of species and geographic imbalance in the training data.

The class balance, both in terms of species and locations, is an important limitation for this work. The train/val/test splits in iNatSounds were constructed to mitigate these imbalances to some extent. The number of audio recordings per species is capped at 1000 for the training set. This sub-sampling is done as follows: first they are clustered geographically and then recordings are sampled at random from clusters in a round-robin fashion to increase geographic diversity. Despite this, there is still geographical bias in our model (Fig 3, right).

We additionally explore resampling strategies to mitigate this species and geographic imbalance. For each recording of a given species at a particular location, we compute the number of training recordings either (i) of that species or (ii) in the same h3 hexagon. With this count $n$, we assign the recording a a weight inversely proportional to $n$ ($w \propto \frac{1}{n}$) or its square root ($w \propto \frac{1}{\sqrt{n}}$). Performing a weighted random sampling with these weights allows us to balance the species and geographic distributions, respectively. 

We see worse overall performance for each of these resampling strategies, with significant drops in region performance. Location based resampling, with a weight inversely proportional to the square root of the count performs best amongst the different reweighting strategies. We also look at the geographic distribution of performance for these strategies in Fig~\ref{fig:class_balancing}. We see improvements in a few regions in east Europe and east Asia, but changes in the global distribution of performance are not that obvious.

\begin{table}[ht]
    \centering
    \caption{\textbf{Data Resampling.}  For each recording of a given species at a particular location, we compute the number of training recordings either (i) of that species or (ii) in the same h3 hexagon. With this count $n$, we assign the recording a a weight inversely proportional to $n$ ($w \propto \frac{1}{n}$) or its square root ($w \propto \frac{1}{\sqrt{n}}$). Performing a weighted random sampling with these weights allows us to balance the species and geographic distributions, respectively.}
    \label{tab:resampling}
    \begin{tabular}{lccccc}
    \toprule
    Method & Median Error & City & Region & Country & Continent \\
    \midrule
    no resampling & 1082 & 6.4 & 17.2 & 41.0 & 71.2 \\
    \midrule
    \rowcolor{tabledarkgray}\multicolumn{6}{l}{$w \propto \frac{1}{n}$} \\
    \midrule
    species & 1651 & 4.3 & 12.0 & 31.2 & 59.2 \\
    location & 1601 & 1.7 & 6.5 & 26.3 & 61.9 \\
    \midrule
    \rowcolor{tabledarkgray}\multicolumn{6}{l}{$w \propto \frac{1}{\sqrt{n}}$} \\
    \midrule
    species & 1190 & 5.7 & 15.6 & 38.9 & 67.8 \\
    location & 1204 & 4.6 & 13.6 & 37.4 & 69.2 \\
    location $\times$ species & 1440 & 3.2 & 10.6 & 31.7 & 64.7 \\
    \bottomrule
    \end{tabular}
\end{table}

\subsection{Time in geolocation}
\label{sec:supp:time_modality}

Please see Table~\ref{tab:month} for experiments incorporating time into audio geolocation as an auxiliary training objective. 
Time is an important modality for bioacoustic geolocation because of seasonal species migrations. For a simple way of integrating time into our approach, we explore using the month of the recording as an auxiliary metadata loss. We model month prediction as 12-way classification, including an additional prediction head and corresponding cross entropy loss. 

These results indicate only marginal or no improvements upon the addition of predicting month as an auxiliary task. We still believe time to be an important factor for geolocation, but leave further explorations in this direction for future work.

\begin{table}[!h]
    \centering
    \caption{\textbf{Incorporating time into audio geolocation.} Experiments with month-of-recording prediction as an auxiliary training objective. We experiment with a few different weights on the month prediction loss. }
    \label{tab:month}
    \begin{tabular}{lccccc}
         \toprule
Method & Median Error & City & Region & Country & Continent \\
\midrule
AG-CLIP & 1082 & 06.4 & 17.2 & 41.0 & 71.2 \\
\midrule
AG-CLIP + month 0.8 & 1074 & 06.2 & 16.8 & 41.2 & 71.0 \\
AG-CLIP + month 0.1 & 1113 & 06.2 & 16.6 & 40.3 & 70.5 \\
AG-CLIP + month 1.0 & 1082 & 05.8 & 16.4 & 41.0 & 71.0 \\
AG-CLIP + month 2.0 & 1122 & 05.1 & 15.0 & 40.1 & 70.5 \\
AG-CLIP + month 5.0 & 1282 & 04.3 & 12.8 & 36.4 & 67.2 \\
\bottomrule
    \end{tabular}
\end{table}

\subsection{Embedding Visualizations}
\label{sec:supp:embeddings}
In Table~\ref{tab:pretrain}, we test whether features learnt for geolocation transfer well to the task of species identification. First, we (pre)train a model for audio geolocation from scratch. Then, we (post)train this model for species identification. We compare this with a model trained for species identification from scratch.
These results suggest that the features learnt geolocation models are helpful for species identification.

\begin{figure}[ht]
    \centering
    \includegraphics[width=\linewidth]{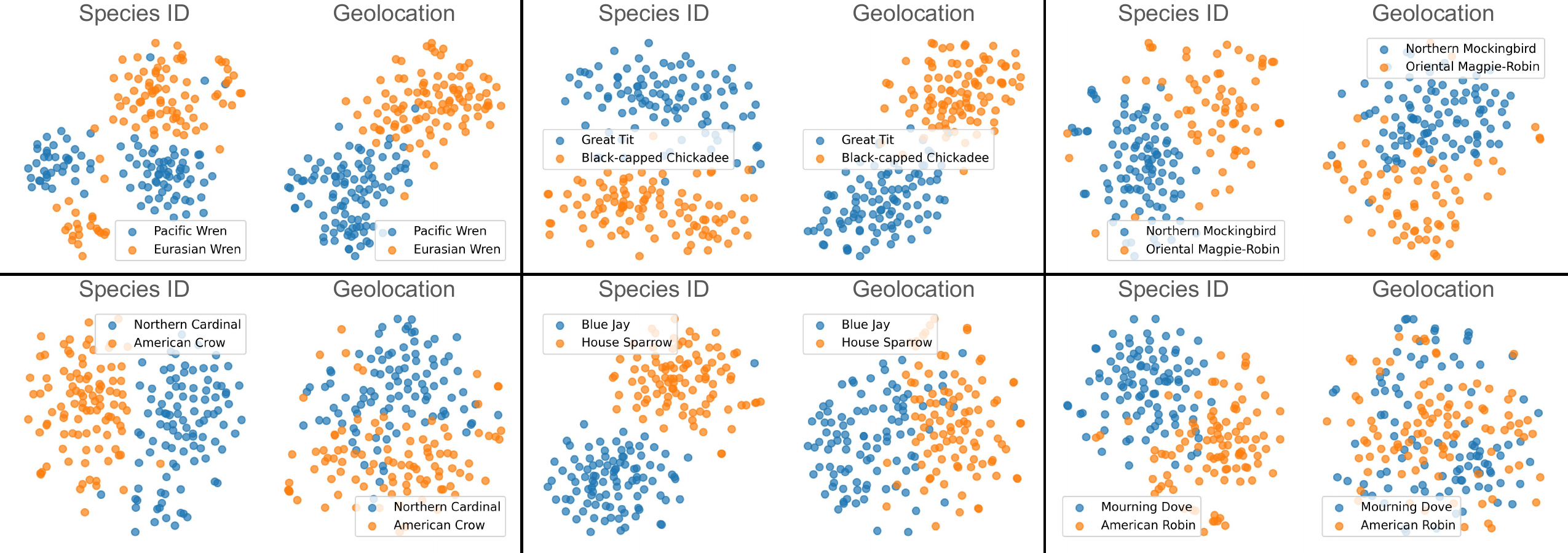}
    \caption{\textbf{Utility of features for species identification.} We visualize t-SNE projections of learned embeddings of species-ID and audio geolocation models for a few species pairs. The first row includes species pairs that tend to sound similar but are found geographically distant regions, while the second row includes species pairs that sound different but are found in the same regions.}
    \label{fig:tsne_species}
\end{figure}

To further evaluate the embeddings directly, we obtain t-SNE plots of particular species pairs and compare them for embeddings learnt by geolocation and species identification models. Please see these plots in Fig~\ref{fig:tsne_species}. These plots suggest that geolocation models are:
\begin{enumerate}
    \item Good at differentiating species that sound similar but are from different geographic regions
    \item Bad at differentiating between species that sound distinct but are from similar geographic regions
\end{enumerate}

\begin{figure}[ht]
    \centering
    \includegraphics[width=\linewidth]{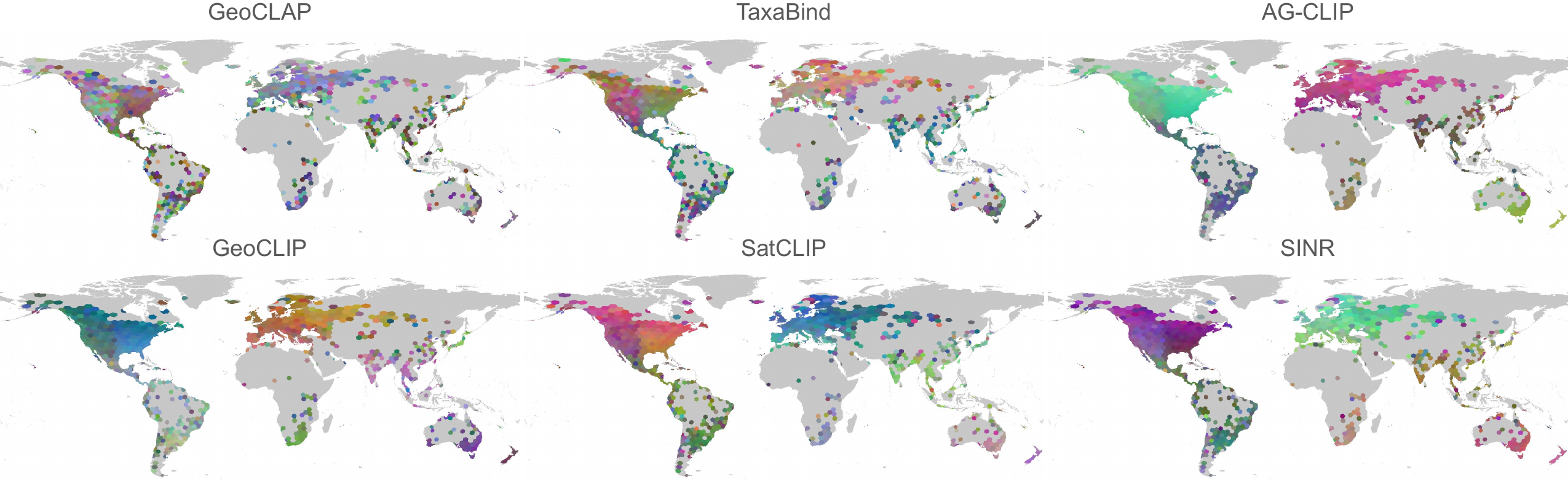}
    \caption{\textbf{Audio Embedding Visualizations.} For each h3 hexagon, we aggregate the embeddings from all test audio within that hexagon and visualize the 3D t-SNE projection of this aggregate embedding as RGB values. The colors are not comparable across different plots, but within a plot, similar colors can be interpreted as similar embeddings. }
    \label{fig:tsne_locations}
\end{figure}

Please see Fig~\ref{fig:tsne_locations} for additional visualization of audio embeddings. For each h3 hexagon, we aggregate the embeddings from all test audio within that hexagon and visualize the 3D t-SNE projection of this aggregate embedding as RGB values. The colors are not comparable across different plots, but within a plot, similar colors can be interpreted as similar embeddings.
For the baselines Taxabind and GeoCLAP, the embeddings seem somewhat noisy, sometimes having significantly different embeddings for neighboring regions. For Taxabind, we see some consistency for the eastern US. Our audio embeddings, with SatCLIP, SINR or GeoCLIP location encoders seem much more uniform, with noticeable global patterns.

\begin{table}[!h]
    \centering
    \caption{\textbf{Utility for species identification.} We explore the use of audio geolocation as a pretraining strategy for species idenficiation. }
    \label{tab:pretrain}
    \begin{tabular}{lcc}
         \toprule
        Experiment & Top 1 & Top 5 \\
        \midrule
        Train from scratch & 48.4 & 70.2 \\
        Pretrain with Geolocation & 50.7 & 71.6 \\
    \bottomrule
    \end{tabular}
\end{table}

\subsection{Geolocation performance with uncertainty in SINR checklists}
\label{sec:supp:sinr_uncertain}

 To quantify the uncertainty of SINR checklists, we compute the entropy of the predicted class-wise probabilities, normalized by the number of predicted positives. We then iteratively drop the most uncertain recordings (with highest entropy) from our evaluation and plot the median geolocation error as a function of the fraction of data removed.

\begin{figure}[ht]
    \centering
    \includegraphics[width=0.5\linewidth]{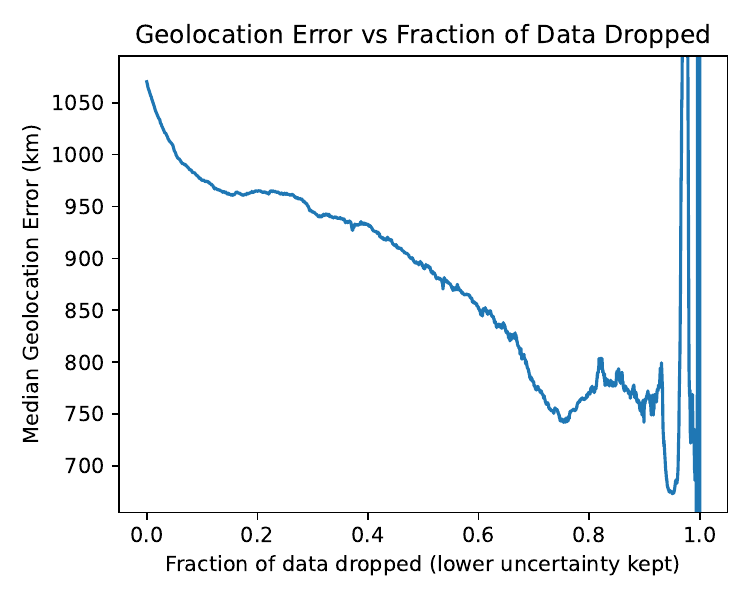}
    \caption{\textbf{Geolocation error and SINR uncertainty.} To capture the uncertainty of SINR checklists, we compute the entropy of the predicted class-wise probabilities, normalized by the number of predicted positives. We then drop the most uncertain recordings from our evaluation and plot the median geolocation errors as a function of the fraction of data dropped. }
    \label{fig:geo_vs_sinr}
\end{figure}

Please see Fig~\ref{fig:geo_vs_sinr} for a comparison of the geolocation error with the uncertainty of predicting  SINR checklists. The plot indicates that geolocation performance tends to improve with lower uncertainty in SINR checklist predictions. For the full dataset, the median geolocation error is 1082 km, which drops to around 760 km for the top 10\% of recordings with the lowest SINR uncertainty. The plot is quite noisy towards the right end (very small fractions of data left), reflecting exceptions to this trend with poor geolocation performance despite highly confident SINR predictions. Broadly, lower uncertainty in the SINR checklists likely corresponds to regions with more distinctive species compositions and this trend suggests that such regions are easier to geolocate.

\subsection{Compute Resources}\label{sec:supp:compute}

We used a single A100 GPU with 80 GB GPU memory for all experiments. On a node with 1 GPU and 8 CPUs, one experiment of AG-CLIP takes 2.5 hours on iNatSounds (training + testing) and about 20 minutes for XCDC (testing). Other approaches of species oracles, naive baselines and species ranges are much faster. Classification and regression take about the same training time as AG-CLIP, and are faster at inference since retrieval methods need location galleries for prediction.

\end{document}